\documentclass[fleqn,usenatbib]{mnras}

\usepackage{graphicx}
\usepackage{amsmath,amssymb}
\usepackage{pdflscape}
\usepackage{newtxtext,newtxmath}

\usepackage[T1]{fontenc}

\DeclareRobustCommand{\VAN}[3]{#2}
\let\VANthebibliography\thebibliography
\def\thebibliography{\DeclareRobustCommand{\VAN}[3]{##3}\VANthebibliography}

\title[Protocluster candidates at $z\sim3\mathrm{-}5$]{An enhanced abundance of bright galaxies in protocluster candidates at $\mathbf{z\sim3\mathrm{-}5}$}

\author[J. Toshikawa et al.]{
Jun Toshikawa,$^{1,2}$\thanks{E-mail: jt2155@bath.ac.uk} Stijn Wuyts,$^1$ Nobunari Kashikawa,$^3$ Chengze Liu,$^4$ Marcin Sawicki,$^5$
\newauthor Roderik Overzier,$^6$ Mariko Kubo,$^7$ Hisakazu Uchiyama,$^8$ Kei Ito,$^3$ Malcolm Bremer,$^9$ Yoshiaki Ono,$^{10}$
\newauthor Tadayuki Kodama,$^7$ Yen-Ting Lin$,^{11}$ Tomoki Saito$^2$\\
$^1$Department of Physics, University of Bath, Claverton Down, Bath, BA2 7AY, UK\\
$^2$Nishi-Harima Astronomical Observatory, Center for Astronomy, University of Hyogo, Sayo, Hyogo 679-5313, Japan\\
$^3$Department of Astronomy, University of Tokyo, Hongo, Tokyo 113-0033, Japan\\
$^4$Department of Astronomy, School of Physics and Astronomy, and Shanghai Key Laboratory for Particle Physics and Cosmology, Shanghai Jiao Tong\\University, Shanghai 200240, China\\
$^5$Institute for Computational Astrophysics and Department of Astronomy and Physics, Saint Mary's University, 923 Robie Street, Halifax, Nova Scotia,\\B3H 3C3, Canada\\
$^6$Observat\'orio Nacional, Rua Jos\'e Cristino, 77. CEP 20921-400, S\~ao Crist\'ov\~ao, Rio de Janeiro-RJ, Brazil\\
$^7$Astronomical Institute, Tohoku University, 6-3, Aramaki, Aoba, Sendai, Miyagi, 980-8578, Japan\\
$^8$Research Center for Space and Cosmic Evolution, Ehime University, 2-5 Bunkyo-cho, Matsuyama, Ehime 790-8577, Japan\\
$^9$H H Wills Physics Laboratory, University of Bristol, Tyndall Avenue, Bristol, BS8 1TL, UK\\
$^{10}$Institute for Cosmic Ray Research, The University of Tokyo, 5-1-5 Kashiwa-no-Ha, Kashiwa, Chiba 277- 8582, Japan\\
$^{11}$Institute of Astronomy and Astrophysics, Academia Sinica (ASIAA), Taipei 10617, Taiwan
}

\date{Accepted XXX. Received YYY; in original form ZZZ}

\pubyear{2022}

\begin{document}
\label{firstpage}
\pagerange{\pageref{firstpage}--\pageref{lastpage}}
\maketitle

\begin{abstract}
We present a protocluster search covering $z\sim3$ to $z\sim5$ based on the combination of the Hyper SuprimeCam
Subaru Strategic Programme and the CFHT Large Area $U$-band Deep Survey.
We identify about 30 protocluster candidates per unit redshift over the $\sim25\,\mathrm{deg^2}$ area of the
Deep/Ultra-Deep layer.
Protocluster candidates are selected as regions with a significantly enhanced surface density of dropout galaxies.
With this large sample, we characterise the properties of their individual member galaxies.
We compare the number counts of dropout galaxies in protocluster candidates with that of coeval field galaxies.
Rest-frame UV bright galaxies are over-abundant in protocluster candidates, a trend seen across the full redshift
range studied.
We do not find evidence for their spatial distribution within protocluster candidates to be distinct from their
fainter counterparts, nor for their UV colour to be different from that of field galaxies with the same brightness.
Cosmological simulations predict this bright-end excess, with the main cause being a richer population of massive
galaxies, with only a minor contribution from an enhancement in star formation activity (and therefore UV
emission) at fixed mass.
$U$-to-$K$ SED modelling of our observed samples supports this interpretation.
This environmental differentiation in number counts is already in place at $z\sim5$, with no significant redshift
dependence over the range in lookback times probed.
These observational results and model predictions suggest that the cosmic clock is ahead in high-density
environments.
\end{abstract}

\begin{keywords}
galaxies: evolution -- galaxies: high-redshift
\end{keywords}

\section{Introduction}
The exploration of the growth and evolution of the largest cosmic structures and how this growth influences their
galaxy populations are the central drivers of current astrophysical and cosmological research.
Galaxy clusters act as key laboratories for this research for two main reasons.
Firstly, there are clear differences between the cluster and field galaxy populations, presumably due to the
different evolutionary histories of their environments \citep[e.g.,][]{dressler80,thomas05,trudeau22}.
Secondly, galaxy clusters are located at the nodes of the cosmic web and are therefore sensitive probes of the
growth of this cosmic structure \citep[e.g.,][]{peebles80,alpaslan14}.
Consequently, galaxy clusters are a crucial bridge between galaxy evolution and the growth of cosmic structures,
and equivalently between astrophysical and cosmological phenomena.

To reveal details of the formation and early evolution of galaxy clusters, it is necessary to directly observe
their early growth phase, so-called ``protoclusters'', at high redshifts.
The first clear examples of protoclusters were discovered at $z\sim2\mathrm{-}3$
\citep{steidel98,pascarelle98,pentericci00}.
Subsequently, in order to expand protocluster searches, distant radio galaxies (RGs) or quasars (QSOs) were often
used as the signposts of galaxy overdensities because such galaxies should be formed in massive halos
\citep[e.g.,][]{venemans07}.
Protocluster searches around RGs were significantly enlarged by \citet{wylezalek13} using a large sample of
$\sim400$ RGs at $z\sim1\mathrm{-}3$, followed by the confirmation of 16 structures through follow-up spectroscopy
\citep{noirot18}.
Although these works clearly demonstrate that many powerful distant RGs reside in high-density environments or
protoclusters, the physical mechanisms connecting RGs to protoclusters are still unclear
\citep{hatch14,marinello20}.

Contemporaneously, the advent of sensitive, comparatively wide-field imaging facilities and surveys allowed
systematic protocluster searches without using other objects as signposts, the large areas/volumes being required
because of the expected rarity of protoclusters.
The 8-m Subaru telescope, with the optical imaging instrument of SuprimeCam at its prime focus, enabled us to find
protoclusters without preselection by RGs or QSOs \citep{shimasaku03,ouchi05,toshikawa12}.
Other public surveys were also used to search for protoclusters: for example, \citet{toshikawa16} identified 21
protocluster candidates as the surface overdense regions of dropout galaxies at $z\sim3\mathrm{-}6$ in the
$4\,\mathrm{deg^2}$ area of the Canada-France-Hawaii Telescope (CFHT) Legacy Survey Deep Fields, and
\citet{chiang14} constructed a catalogue of 36 protocluster candidates at $z\sim1.5\mathrm{-}3$ over a
$1.6\,\mathrm{deg^2}$ area by exploiting photometric redshifts estimated by multi-wavelength data of the
COSMOS/UltraVISTA survey.
Furthermore, spectroscopic programmes like the VIMOS Ultra-Deep survey discovered several protoclusters based on
the direct measurement of three-dimensional galaxy number density \citep{cucciati18,lemaux18}.
These works demonstrated that blank surveys are essential to construct more complete samples of protoclusters
though searches only around RGs may be effective to find some of them.

Combining various surveys and methods, the sample size as well as redshift range of protoclusters
is increasing \citep[e.g.,][]{hu21}.
Especially, the James Webb Space Telescope (JWST) enables to discover protoclusters even at $z\gtrsim7$ by its high
sensitivity at near-infrared wavelength.
\citet{morishita22} confirmed a protocluster at $z=7.9$, and \citet{laporte22} reported a candidate at $z=7.7$.
However, it should be noted that the completeness of protocluster samples is still very small.
Some protoclusters grow up more rapidly at later epochs, implying that they may not hold sufficient member
galaxies to be identifiable at an early epoch.
On top of this inherent cause of incompleteness, projection effects dilute overdensity signals from protoclusters
as illustrated by Figure~13 of \citet{chiang13}.
\citet{toshikawa16} also predict that the completeness of protoclusters selected from dropout galaxies is only
at the $\lesssim10\%$ level based on model comparisons.
In actual observations, \citet{noirot18} and \citet{toshikawa20} found that some surface overdense regions are
composed of several structures along the line-of-sight direction.

Beyond just identifications, the galaxy properties of individual protoclusters at $z\sim2\mathrm{-}3$ are well
studied from various viewpoints.
Massive quiescent galaxies already appear in high-density environments at $z\sim3$, while star-forming galaxies
are still dominant in protoclusters \citep[e.g.,][]{wang16,kubo21}.
There is diversity too, as illustrated by two protoclusters at $z\sim3$ which show different galaxy populations
from each other though they were identified from the same dataset and via the same method \citep{shi20,shi21}.
As for active galactic nucleus (AGN) activity, there are both examples of protoclusters which host more and less
AGNs than coeval field environments \citep[e.g.,][]{lehmer13,macuga19,vito20}.
Similarly, protoclusters exhibit controversial results regarding their member galaxies' metallicity
\citep[e.g.,][]{shimakawa15,chartab21,sattari21}.
These diverse conditions of protoclusters may reflect distinct developmental stages of cluster formation.
However, it is difficult to lay out the general picture of cluster formation or galaxy evolution in high-density
environments based on a random collection of protoclusters.
In order to develop an understanding of the relation between galaxy evolution and the formation of cosmic
structures, a systematic sample of protoclusters across cosmic time is imperative.

The wide-field imaging capability of the Subaru telescope was significantly improved with the installation of
Hyper SuprimeCam (HSC), enabling the Subaru Strategic Programme with HSC (HSC-SSP) which has been ongoing since
2014 \citep{aihara18}.
The HSC-SSP is composed of three layers: Wide ($\sim1000\,\mathrm{deg^2}$, $i$-band depth of
$m_i\sim26.0\,\mathrm{mag}$), Deep ($\sim26\,\mathrm{deg^2}$, $m_i\sim26.5\,\mathrm{mag}$), and UltraDeep
($\sim3\,\mathrm{deg^2}$, $m_i\sim27.0\,\mathrm{mag}$).
We have carried out a protocluster search at $z\sim4$ over $>100\,\mathrm{deg^2}$ area by using the Wide layer of
the 1st-year HSC-SSP data release \citep{toshikawa18}.
This has resulted in a systematic sample of $>100$ protocluster candidates and enabled us, for the
first time, to estimate the dark matter halo mass of protoclusters from clustering analysis.
The derived relation between number density and correlation length of protocluster candidates at $z\sim4$ was
compared with the prediction of a $\Lambda$CDM model.
In addition, based on this systematic sample, we have conducted an investigation of QSO environments
\citep{onoue18,uchiyama18}, bright protocluster galaxies \citep{ito19,ito20}, and infrared emission from
UV-selected protoclusters \citep{kubo19}.
Another unique feature of the HSC-SSP is a survey component observing through multiple narrow-band filters.
This enables the selection of galaxies within a thin redshift slice.
In particular, the HSC-SSP exploits narrow-band filters corresponding to the wavelengths of Ly$\alpha$ emissions
at $z=5.7$ and 6.6; thus, \citet{higuchi19} explored the implications of early clustering on cosmic reionisation
on top of the identifications of 40 protocluster candidates, and some of them were spectroscopically confirmed by
\citet{harikane19}.

This paper expands our protocluster search to a wider redshift range by making use of the Deep and UltraDeep (DUD)
layer of the latest HSC-SSP data release.
In the DUD layer, unlike the Wide layer, $U$-band imaging data are also provided by the CFHT Large Area $U$-band
Deep Survey\footnote{\url{https://www.clauds.net}} \citep[CLAUDS;][]{sawicki19}.
This deep multi-wavelength dataset allows us to identify sufficiently large numbers of high-redshift galaxies to
map cosmic structures from low- to high-density regions, thus probing environmental densities ranging from those
of voids to those of protoclusters.
We have identified $\ga30$ protocluster candidates at each unit redshift between $z\sim3$ and $z\sim5$, and
provide their key properties as part of this paper.
By the combination of a large sample and deep multi-wavelength dataset, we address the environmental differences
of galaxy properties from $z\sim5$ to $z\sim3$, and investigate the physical nature of protocluster candidates
more closely in the context of cosmological simulations.

The structure of this paper is as follows.
Section~\ref{sec:data} describes the imaging data, the selection of high-redshift galaxies, and how protocluster
candidates are identified.
In Section~\ref{sec:model}, the same method to search for protoclusters is applied to light-cone models generated
from a cosmological simulation to understand protocluster properties imprinted in observations.
We explore possible differences between protocluster and field galaxies based on the number counts of high-redshift
galaxies, the physical properties derived by spectral energy distribution (SED) modelling, and the spatial
distribution of protocluster members within a protocluster candidate in Section~\ref{sec:res}.
The derived results are compared with theoretical predictions and other observations of protoclusters in order to
discuss the physical properties of protocluster galaxies in Section~\ref{sec:dis}.
The conclusions are provided in Section~\ref{sec:con}.
The following cosmological parameters are assumed: $\Omega_\mathrm{M}=0.3$, $\Omega_\mathrm{\Lambda}=0.7$,
$H_0=70\,\mathrm{km\,s^{-1}\,Mpc^{-1}}$.  Magnitudes are given in the AB system.

\section{Data \& protocluster candidates} \label{sec:data}
\begin{table*}
\caption{Photometric data of the HSC-SSP and CLAUDS}
\label{tab:images}
\begin{tabular}{ccccccccccc}
\hline
Name & R.A. & Decl. & Area$^\text{a}$ & $U$ & $g^\text{b}$ & $r^\text{b}$ & $i^\text{b}$ & $z^\text{b}$ & $y^\text{b}$ \\
 & (J2000) & (J2000) & ($\mathrm{deg^2}$) & (mag) & (mag) & (mag) & (mag) & (mag) & (mag) \\
\hline
COSMOS & 10:00:24 & $+$02:12:39 & 6.55 (4.67) & 27.6 & 26.7 (27.4) & 26.4 (27.1) & 26.3 (27.0) & 25.8 (26.5) & 24.9 (25.8) \\
DEEP2-3 & 23:28:18 & $-$00:15:58 & 6.33 (3.91) & 27.4 & 26.6 & 26.3 & 25.9 & 25.5 & 24.6 \\
ELAIS-N1 & 16:10:56 & $+$54:58:14 & 5.71 (3.90) & 27.4 & 26.8 & 26.3 & 26.0 & 25.5 & 24.5 \\
XMM-LSS & 02:18:23 & $-$04:52:52 & 6.65 (4.84) & 27.6 & 26.7 (27.1) & 26.2 (26.8) & 25.9 (26.5) & 25.6 (26.1) & 24.5 (24.5) \\
\hline
\end{tabular}
\\
\footnotesize{$^\text{a}$The coverage of CLAUDS is given in parentheses.\\
    $^\text{b}$The depth of the Ultradeep area is given in parentheses.}
\end{table*}

\subsection{$\mathbf{z\sim3\mathrm{-}5}$ galaxies in the HSC-SSP \& CLAUDS}
We use optical multi-band imaging data from $g$-band to $y$-band in the DUD layer of the HSC-SSP together with
the $U$-band\footnote{COSMOS, DEEP2-3, and ELAIS-N1 were newly observed by CLAUDS with the $u$-band of CFHT.
Archival data taken with the $u^\ast$-band existed in parts of the XMM-LSS and these data were used along with new
$u^\ast$-band observations in that field.
Additionally, the central region of the COSMOS field had deep $u^\ast$-band archival data and was also observed
with $u$-band, so that $\sim1\mathrm{deg^2}$ region has imaging in both $u$ and $u^\ast$ ($u$- and $u^\ast$-bands
are kept separate).
Thus, the CLAUDS dataset is composed of both $u$- and $u^\ast$-bands, and we refer to $u$- and/or $u^\ast$-bands
as $U$-band following the notation in \citet{sawicki19}.}
data of CLAUDS to identify protocluster candidates
marked out by overdensities of dropout galaxies.
The S21A HSC-SSP data release is used in this study.
The DUD layer is composed of four independent fields: COSMOS, DEEP2-3, ELAIS-N1, and XMM-LSS.
The $i$-band $5\sigma$ depth of the DUD layer is $26.5\,\mathrm{mag}$ or deeper over $25.2\,\mathrm{deg^2}$,
though the coverage of the CLAUDS $U$-band imaging is slightly smaller ($17.3\,\mathrm{deg^2}$) than that of the
$g$-to-$y$ coverage of the HSC-SSP.
Table~\ref{tab:images} summarises the effective area and $5\sigma$ depth of each field.
We refer the reader to \citet{aihara22} for more details on the image reduction, object detection, and photometry.
In any event, the combined HSC-SSP plus CLAUDS dataset is deep enough to select galaxies down to the characteristic
magnitude at $z\sim3\mathrm{-}5$ via the Lyman break technique, tracing the redshifted Lyman-limit at
$\lambda_{\rm rest} = 912$\AA\ by means of appropriate colour-colour cuts \citep[see, e.g.,][]{steidel92}.

\begin{figure*}
\includegraphics[scale=0.43,bb=10 0 1180 400]{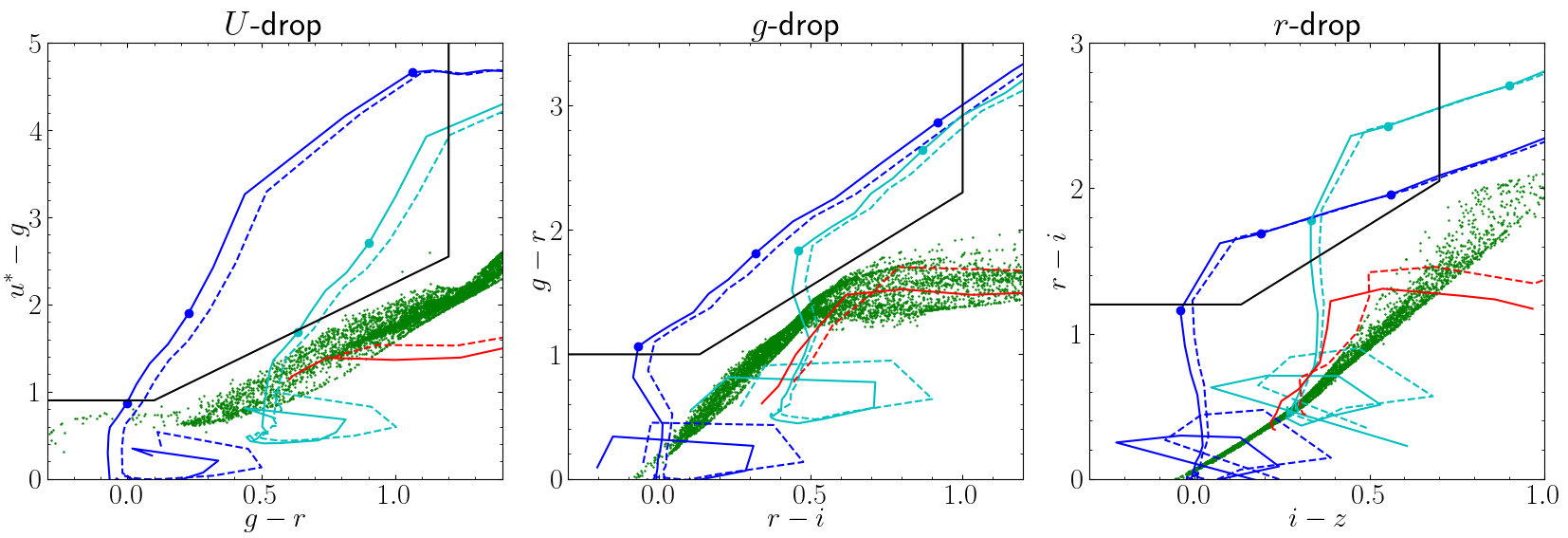}
\caption{Demonstration of dropout galaxy selection on two-color diagrams.
    The thick black lines show the borders of our dropout galaxy selection window.
    The solid blue lines indicate the observed-frame colours of a young star-forming galaxy template
    (constant star-formation-history model with $\mathrm{age}=200\,\mathrm{Myr}$, a \citet{salpeter55} initial
    mass function, stellar metallicity of $0.2Z_{\sun}$, and dust attenuation of $E(B-V)=0.1$), computed using
    CIGALE and shifted in redshift space \citep{meiksin06}. Dashed lines show the equivalent tracks for a
    star-forming template with $\mathrm{age}=600\,\mathrm{Myr}$.
    The redshift track of a dustier ($E(B-V)=0.5$) star-forming galaxy template is shown in cyan.
    On each solid curve, three markers correspond to $z=2.8,\,3.2,\,3.6$, $z=3.6,\,4.2,\,4.6$, and
    $z=4.7,\,5.0,\,5.3$ for $U$-, $g$-, and $r$-dropout galaxies, respectively .
    Two red lines mark the colour evolution with redshift (over the range $z=0\mathrm{-}1.5$) for quiescent galaxy
    templates with ages of 2 (solid) and $5\,\mathrm{Gyr}$ (dashed).  Green dots are dwarf stars from the TRILEGAL
    Galactic model \citep{girardi05}.}
\label{fig:2cd}
\end{figure*}

From the photometric catalogue of the HSC-SSP in the DUD layer, $g$- and $r$-dropout galaxies are selected
following the colour criteria depicted in Figure~\ref{fig:2cd}:
\begin{description}
\item[$g$-dropouts:] $1.0<(g-r) \wedge -1.0<(r-i)<1.0 \wedge 1.5(r-i)<(g-r)-0.8$,
\item[$r$-dropouts:] $1.2<(r-i) \wedge -1.0<(i-z)<0.7 \wedge 1.5(i-z)<(r-i)-1.0$.
\end{description}
For $U$-dropout galaxies, the colour selection criteria are:
\begin{eqnarray*}
0.9<(u^*-g) \wedge (g-r)<1.2 \wedge 1.5(g-r)<(u^*-g)-0.75, \>\mathrm{or}\\
0.98<(u-g) \wedge (g-r)<1.2 \wedge 1.99(g-r)<(u^*-g)-0.68.
\end{eqnarray*}
The adjustment of colour selection criteria for $U$-dropout galaxies depending on the available $U$-band filter is
determined by the locus of stars in the $ugr$ and $u^*gr$ diagrams.
The typical redshift windows of these colour criteria of dropout selection are $z\sim2.6\mathrm{-}3.6$,
$z\sim3.4\mathrm{-}4.4$, and $z\sim4.3\mathrm{-}5.3$ for $U$-, $g$-, and $r$-dropout galaxies, respectively.
Further details on the dropout selection in the HSC-SSP and CLAUDS and the characteristic properties of the
selected galaxy population are outlined in \citet{ono18} and \citet{harikane22} for $g$- and $r$-dropout galaxies,
and in C. Liu et al. (2023, in preparation) for $U$-dropout galaxies.
Very similar colour selections to identify high-redshift galaxies are commonly used in various studies
\citep[e.g.,][]{burg10,bielby13,bouwens22}.
As shown by \citet{harikane22}, the contamination rate in our dropout samples is expected to be $\lesssim20\%$.
We applied two additional criteria to reduce contamination by low-redshift interlopers.
The first one considers photometric redshifts estimated by the MIZUKI code \citep{tanaka15,tanaka18}, specifically
the upper bound on the central 95\% confidence interval of the photometric redshift probability distribution
($z_\mathrm{phot,95\%}$).
The objects satisfying $z_\mathrm{phot,95\%}<2.3$, $z_\mathrm{phot,95\%}<2.8$, and $z_\mathrm{phot,95\%}<3.8$ in
the catalogues of $U$-, $g$-, and $r$-dropout galaxies, respectively, are regarded as contaminants.
This is more conservative than \citet{harikane22} who imposed a criterion of $z_\mathrm{phot,95\%}<2.0$ for all
dropout samples.
Our thresholds correspond to the lower side of each dropout sample's redshift window minus 0.5.
Secondly, we ignore in this study any areas where the point spread function (PSF) is wrongly measured due to
instrumental or software issues.
The rationale for this second criterion is that an incorrect PSF characterisation can lead to an unrealistic
aperture correction and consequently generate artificial dropout galaxies.
Although the area affected is very small compared to the whole survey area, the density of dropout galaxies in
such regions can be artificially increased through this error in the photometry.
The occasional pipeline failure in measuring the PSF could thus have a significant impact on our protocluster
search if not accounted for.
Through these two additional criteria, $\sim5\%$ of $U$-, $g$-, and $r$-dropout galaxies in the initial catalogues
were removed.

\begin{figure}
\includegraphics[width=\columnwidth,bb=0 0 461 346]{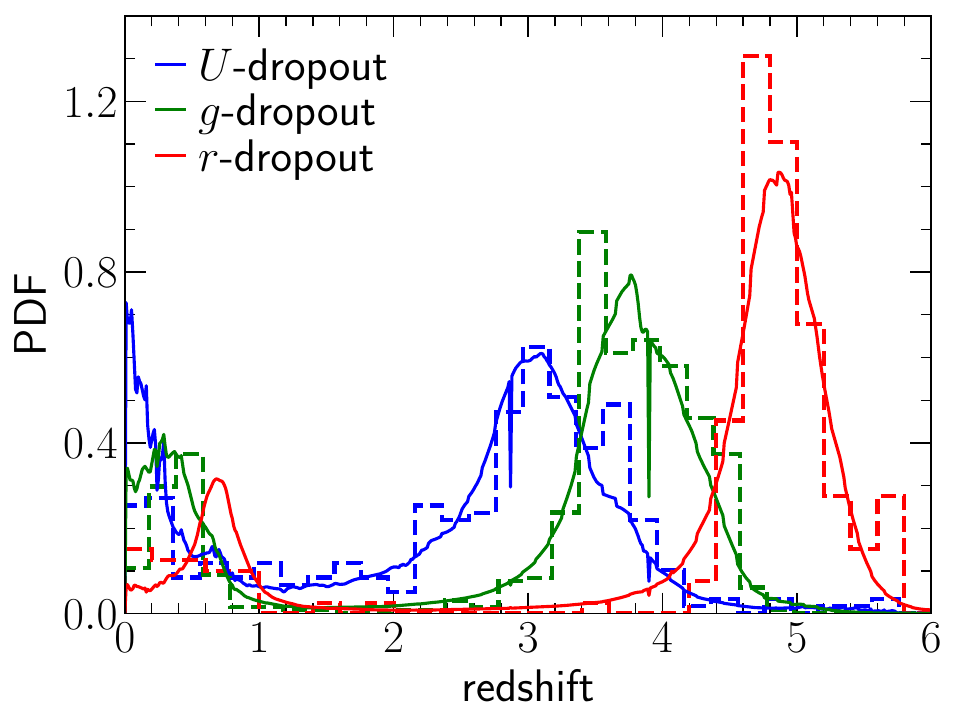}
\caption{Average PDF of phot-$z$ (solid lines) and normalised distribution of spec-$z$ (dashed histogram) for
    dropout galaxies.}
\label{fig:zPDF}
\end{figure}

We evaluate the redshift window and contamination rate of our dropout selection by comparing with photometric
redshifts (phot-$z$) from the COSMOS2020 catalogue by \citet{weaver22}, which covers 6\% of the survey volume
explored in our study.
To do so, we first cross-match our HSC-selected dropout galaxies with sources in the CLASSIC version of the
COSMOS2020 catalogue, within a search radius of $0\farcs5$.
We then compute the average probability distribution function (PDF) of phot-$z$ for each of the $U$-, $g$- and
$r$-dropout samples, and display them in Figure~\ref{fig:zPDF}.\footnote{Internal variation between the different
phot-$z$ catalogues released as part of the COSMOS2020 project can be evaluated from Figure~\ref{fig:zPDF_4type}
in Appendix~\ref{sec:comp}.}
Strong peaks are clearly present at $z\sim3.1$, 3.8, and 4.8 for $U$-, $g$-, and $r$-dropouts, respectively.
A secondary peak at $z<2$ indicating low-$z$ contaminations is also visible, which is attributed to the well-known
Balmer/Lyman break confusion.
The integrated PDF over $z<2$ is found to be 0.15-0.25, meaning that the probability of low-$z$ contamination in
our dropout samples would be 15\%-25\%.
In addition, we compared our dropout catalogues to spectroscopic redshift (spec-$z$) catalogues from the
literature.
Although the sample size is much smaller (summing to a total of 1,165 redshifts), we similarly derived an estimate
of the low-$z$ contamination rate, arriving at 10\%-20\%.
The smaller contamination rate derived from spec-$z$ catalogues suggests that the COSMOS2020 phot-$z$ catalogue
itself contains a certain fraction of contaminants.
Altogether, these comparisons boost confidence in our selection of dropout galaxies from the HSC-SSP dataset.
It should be noted that we do not remove contaminations identified by the cross-matching to phot-$z$ or spec-$z$
catalogues because the coverage of COSMOS2020 or spectroscopic surveys is much smaller than the DUD layer, which
would result in an inhomogeneous selection.
If contaminations are uniformly distributed over sky, they will not have a significant effect on our protocluster
search.

Based on the resulting catalogues of $U$-, $g$-, and $r$-dropout galaxies, we next measure the surface number
density of dropout galaxies to identify overdense regions as protocluster candidates (Section~\ref{sec:pcl}).
It should be noted that the HSC-SSP dataset enables us to select higher-redshift samples of $i$- or $z$-dropout
galaxies as performed by \citet{harikane22}; however, their resulting sample sizes are too small to accurately map
overdensity across the whole DUD layer.
In a large fraction ($\sim60\%$) of the DUD layer, near-infrared (NIR) images are also provided by the Deep UKIRT
NEar-Infrared Steward Survey (DUNES$^2$) and the archival surveys of the UKIDSS Ultra-Deep Survey
\citep{lawrence07} and UltraVISTA \citep{mccracken12}.
The $5\sigma$ limiting magnitude of $K$- or $K_s$-band varies from field to field
($\sim23.0\mathrm{-}25.3\,\mathrm{mag}$).
Although this NIR dataset is not used for the selection of dropout galaxies or protocluster candidates, we will
make use of it for the estimation of galaxy properties (e.g., stellar mass and star formation rate).

\subsection{Identification of protocluster candidates} \label{sec:pcl}
The local surface number density of dropout galaxies is quantified by counting the number of galaxies within fixed
apertures of $0.75\,\mathrm{proper\>Mpc}$ (pMpc) radius.
For $U$-, $g$-, and $r$-dropout galaxies, this corresponds to angular radii of 1.6, 1.8, and
$1.9\,\mathrm{arcmin}$, respectively.
Semi-analytical galaxy-formation models built on $N$-body dark matter simulations predict roughly $\sim50\%$ of
protocluster galaxies to be enclosed in such an aperture \citep[e.g.,][]{chiang13,muldrew15}.
Since the spatial distribution of protocluster galaxies depends on their descendant halo mass, the fraction of
protocluster galaxies within a fixed aperture is smaller in the case of more massive protoclusters.
Such massive protoclusters are, nevertheless, expected to exhibit a significant overdensity within
$0.75\,\mathrm{pMpc}$ radius of their centres.
Apertures are distributed over the whole DUD layer in a grid pattern at intervals of $0.5\,\mathrm{arcmin}$.
The average and standard deviation, $\sigma$, of galaxy numbers in the apertures of a given field are determined.
Masked areas are assumed to have the average number density, though apertures in which the masked area exceeds
20\% are discarded from our analysis.
We then define overdense regions which reach $>4\sigma$ at their peak as protocluster candidates.
This method of overdensity measurement is identical to that in \citet{toshikawa16}, which resulted in the
successful discovery of several protoclusters by follow-up spectroscopy \citep{toshikawa16,toshikawa20}.

\begin{figure*}
\includegraphics[scale=0.69,bb=10 0 900 570]{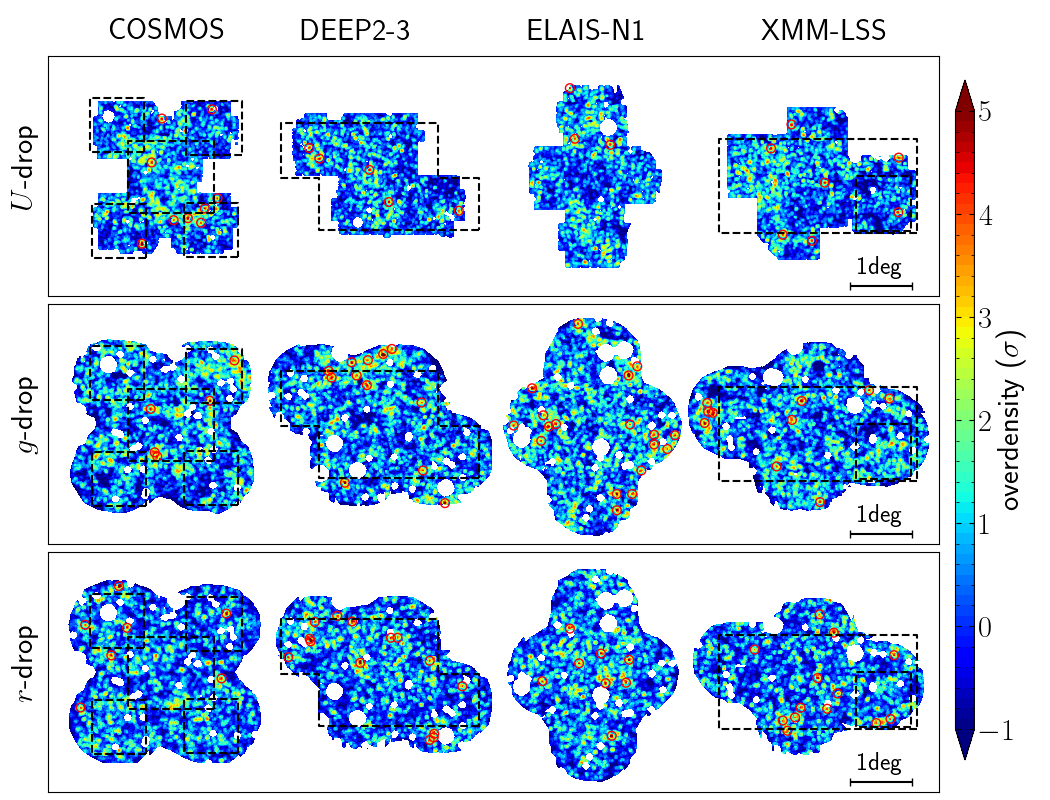}
\caption{Overdensity contours of $U$-, $g$-, and $r$-dropout galaxies in the DUD layer of COSMOS, DEEP2-3,
    ELAIS-N1, and XMM-LSS.
    Higher density regions are indicated by redder colours, and protocluster candidates are marked by red circles.
    The dashed black boxes indicate the footprints of $K(K_s)$-band imaging obtained from UVISTA, UKIDSS/UDS,
    VIDEO, and DUNES$^2$.
    UKIDSS/DXS also covers some part of the HSC-Deep layer but is much shallower than the aforementioned NIR
    surveys.}
\label{fig:cntr}
\end{figure*}

\begin{figure*}
\includegraphics[scale=0.56,bb=0 0 900 440]{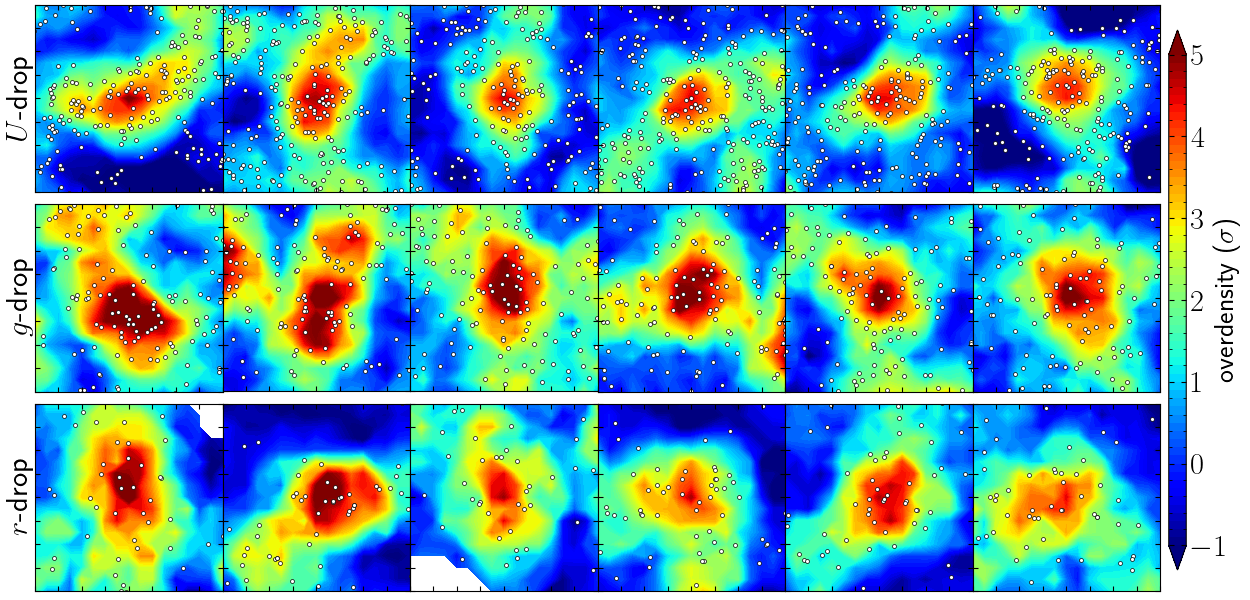}
\caption{Examples of protocluster candidates.
    The top, middle, and bottom rows indicate the protocluster candidates of $U$-dropout ($z\sim3$), $g$-dropout
    ($z\sim4$), and $r$-dropout ($z\sim5$) galaxies, respectively.
    Each panel shows a $8\times8\mathrm{-arcmin^2}$ area centred at the overdensity peak.
    The white points represent dropout galaxies down to the $5\sigma$ limiting magnitude of the individual area,
    but only one forth (half) of $U$-dropout ($g$-dropout) galaxies are plotted in order to make the figure more
    easily readable.}
\label{fig:pcl}
\end{figure*}

Figure~\ref{fig:cntr} shows the overdensity maps of $U$-, $g$-, and $r$-dropout galaxies in the whole DUD layer
(COSMOS, DEEP2-3, ELAIS-N1, and XMM-LSS).
We have identified 24, 45, and 42 protocluster candidates of $U$-, $g$-, and $r$-dropout galaxies, respectively
(Table~\ref{tab:pcl}).
This corresponds to a number density of $\sim1.7\,\mathrm{deg^{-2}}$ for each dropout sample.
Our threshold of a $>4\sigma$ overdensity for selecting protocluster candidates was chosen to reduce the fraction
of artificial candidates (lines-of-sight containing a projected two-dimensional overdensity of dropout galaxies,
but with no clustering along the line-of-sight), favouring sample purity over sample completeness.
According to the model comparison performed by \citet{toshikawa16}, the contamination rate in our sample of
protocluster candidates is anticipated to be $\sim25\%$, but only 5\%--10\% of protoclusters at the redshift
ranges of the dropout selection would be identified by our threshold of surface overdensity.
We update these estimates based on the latest light-cone model of \citet{henriques15} in Section~\ref{sec:model},
obtaining consistent results.
Protocluster candidates exhibit a wide range of shapes at their $2\sigma$ contours.
Some have small, concentrated shapes while others are elongated or even connected with another protocluster
candidate (Figure~\ref{fig:pcl}).
We have confirmed that, within the 6\% of our survey area in overlap with COSMOS2020 \citep{weaver22}, overdensity
maps consistent with those based on $U$-, $g$-, and $r$-dropout galaxies can be obtained using galaxy selections
based on the various COSMOS2020 photometric redshift catalogues (see Appendix~\ref{sec:comp} for details).
Known protoclusters are also found to be located in regions of higher dropout overdensities
(Appendix~\ref{sec:lite}).

\begin{table*}
\caption{Protocluster candidates in the DUD layer}
\label{tab:pcl}
\begin{tabular}{cccccccccc}
\hline
ID & Field & R.A. (J2000) & Decl. (J2000) & overdensity ($\sigma$) & ID & Field & R.A. (J2000) & Decl. (J2000) & overdensity ($\sigma$) \\
\hline
\multicolumn{5}{c}{$U$-dropout ($N_\mathrm{PCL}=24$)} & 32 & ELAIS-N1 & 16:16:51.7 & $+$55:10:37.5 & 4.02 \\
1 & COSMOS & 09:57:01.0 & $+$01:52:02.8 & 4.35 & 33 & ELAIS-N1 & 16:17:05.7 & $+$54:45:37.5 & 4.39 \\
2 & COSMOS & 09:57:23.0 & $+$03:19:02.8 & 4.52 & 34 & ELAIS-N1 & 16:18:08.5 & $+$55:37:07.5 & 5.17 \\
3 & COSMOS & 09:57:51.1 & $+$01:42:02.8 & 4.48 & 35 & ELAIS-N1 & 16:20:14.1 & $+$55:00:37.5 & 4.48 \\
4 & COSMOS & 09:58:05.1 & $+$01:28:02.8 & 4.04 & 36 & XMM-LSS & 02:17:25.6 & $-$04:13:55.9 & 4.12 \\
5 & COSMOS & 09:58:57.1 & $+$01:32:32.8 & 4.79 & 37 & XMM-LSS & 02:18:45.8 & $-$04:05:55.9 & 4.41 \\
6 & COSMOS & 09:59:53.1 & $+$01:30:32.8 & 4.15 & 38 & XMM-LSS & 02:21:58.5 & $-$05:55:25.9 & 4.33 \\
7 & COSMOS & 10:00:39.2 & $+$03:10:02.8 & 4.52 & 39 & XMM-LSS & 02:23:10.7 & $-$04:16:25.9 & 4.93 \\
8 & COSMOS & 10:01:19.2 & $+$02:27:02.8 & 4.41 & 40 & XMM-LSS & 02:23:48.8 & $-$04:34:55.9 & 4.18 \\
9 & COSMOS & 10:01:57.2 & $+$01:07:32.8 & 5.43 & 41 & XMM-LSS & 02:24:49.0 & $-$05:20:25.9 & 4.18 \\
10 & DEEP2-3 & 23:22:55.2 & $-$00:50:39.8 & 4.02 & 42 & XMM-LSS & 02:28:59.8 & $-$04:27:55.9 & 5.83 \\
11 & DEEP2-3 & 23:27:29.2 & $-$00:42:09.8 & 4.63 & 43 & XMM-LSS & 02:29:17.9 & $-$04:26:55.9 & 5.53 \\
12 & DEEP2-3 & 23:28:45.2 & $-$00:10:39.8 & 4.50 & 44 & XMM-LSS & 02:29:21.9 & $-$04:18:25.9 & 4.05 \\
13 & DEEP2-3 & 23:32:01.2 & $+$00:00:20.2 & 5.18 & 45 & XMM-LSS & 02:29:31.9 & $-$04:37:55.9 & 4.81 \\
14 & DEEP2-3 & 23:32:41.2 & $+$00:10:50.2 & 4.25 & \multicolumn{5}{c}{$r$-dropout ($N_\mathrm{PCL}=42$)} \\
15 & ELAIS-N1 & 16:09:14.5 & $+$55:32:37.5 & 4.34 & 1 & COSMOS & 09:56:27.0 & $+$03:11:02.8 & 4.75 \\
16 & ELAIS-N1 & 16:13:15.3 & $+$55:38:37.5 & 4.06 & 2 & COSMOS & 09:56:45.0 & $+$02:07:32.8 & 4.48 \\
17 & ELAIS-N1 & 16:13:50.2 & $+$56:28:07.5 & 4.53 & 3 & COSMOS & 10:02:55.3 & $+$02:57:02.8 & 4.70 \\
18 & XMM-LSS & 02:16:49.4 & $-$04:20:55.9 & 4.37 & 4 & COSMOS & 10:03:27.3 & $+$03:38:02.8 & 5.32 \\
19 & XMM-LSS & 02:16:51.4 & $-$05:14:25.9 & 4.37 & 5 & COSMOS & 10:03:57.3 & $+$02:30:02.8 & 4.79 \\
20 & XMM-LSS & 02:21:40.4 & $-$04:45:25.9 & 4.44 & 6 & COSMOS & 10:05:39.4 & $+$03:00:02.8 & 4.08 \\
21 & XMM-LSS & 02:22:30.6 & $-$05:42:25.9 & 4.38 & 7 & COSMOS & 10:05:57.4 & $+$01:39:02.8 & 4.21 \\
22 & XMM-LSS & 02:23:50.8 & $-$03:48:25.9 & 4.68 & 8 & DEEP2-3 & 23:22:41.2 & $-$00:30:39.8 & 4.28 \\
23 & XMM-LSS & 02:24:24.9 & $-$05:36:25.9 & 4.21 & 9 & DEEP2-3 & 23:24:31.2 & $-$01:17:09.8 & 4.42 \\
24 & XMM-LSS & 02:25:11.1 & $-$04:11:55.9 & 4.56 & 10 & DEEP2-3 & 23:24:33.2 & $-$01:20:39.8 & 4.29 \\
\multicolumn{5}{c}{$g$-dropout ($N_\mathrm{PCL}=45$)} & 11 & DEEP2-3 & 23:24:49.2 & $-$01:23:09.8 & 4.42 \\
1 & COSMOS & 09:55:55.0 & $+$03:16:02.8 & 4.27 & 12 & DEEP2-3 & 23:24:49.2 & $-$00:05:39.8 & 4.37 \\
2 & COSMOS & 09:57:31.0 & $+$02:36:32.8 & 5.13 & 13 & DEEP2-3 & 23:26:55.2 & $+$00:16:50.2 & 4.20 \\
3 & COSMOS & 10:00:57.2 & $+$01:42:02.8 & 4.81 & 14 & DEEP2-3 & 23:27:21.2 & $+$00:16:50.2 & 4.41 \\
4 & COSMOS & 10:01:07.2 & $+$01:46:02.8 & 4.54 & 15 & DEEP2-3 & 23:29:21.2 & $-$00:07:39.8 & 5.64 \\
5 & COSMOS & 10:01:23.2 & $+$02:28:32.8 & 4.46 & 16 & DEEP2-3 & 23:29:53.2 & $+$00:32:50.2 & 4.62 \\
6 & DEEP2-3 & 23:23:49.2 & $-$01:34:09.8 & 5.19 & 17 & DEEP2-3 & 23:30:51.2 & $+$00:38:20.2 & 4.30 \\
7 & DEEP2-3 & 23:25:15.2 & $-$01:02:09.8 & 4.32 & 18 & DEEP2-3 & 23:32:19.2 & $+$00:32:20.2 & 4.20 \\
8 & DEEP2-3 & 23:25:21.2 & $+$00:04:20.2 & 4.01 & 19 & DEEP2-3 & 23:32:35.2 & $+$00:13:20.2 & 5.71 \\
9 & DEEP2-3 & 23:27:17.2 & $+$00:56:50.2 & 4.15 & 20 & DEEP2-3 & 23:32:39.2 & $+$00:15:50.2 & 5.52 \\
10 & DEEP2-3 & 23:27:51.2 & $+$00:51:50.2 & 5.84 & 21 & DEEP2-3 & 23:34:03.2 & $-$00:02:09.8 & 4.31 \\
11 & DEEP2-3 & 23:28:51.2 & $+$00:45:50.2 & 4.00 & 22 & ELAIS-N1 & 16:07:05.3 & $+$55:14:07.5 & 5.76 \\
12 & DEEP2-3 & 23:28:55.2 & $+$00:21:20.2 & 4.48 & 23 & ELAIS-N1 & 16:07:26.3 & $+$54:51:37.5 & 4.54 \\
13 & DEEP2-3 & 23:29:35.2 & $+$00:30:50.2 & 4.09 & 24 & ELAIS-N1 & 16:09:04.0 & $+$53:59:37.5 & 5.79 \\
14 & DEEP2-3 & 23:29:55.2 & $+$00:43:50.2 & 5.27 & 25 & ELAIS-N1 & 16:09:45.9 & $+$54:51:07.5 & 4.85 \\
15 & DEEP2-3 & 23:30:21.2 & $-$01:14:09.8 & 5.53 & 26 & ELAIS-N1 & 16:10:17.3 & $+$55:20:07.5 & 4.75 \\
16 & DEEP2-3 & 23:31:13.2 & $+$00:28:50.2 & 5.28 & 27 & ELAIS-N1 & 16:12:47.4 & $+$55:10:37.5 & 4.67 \\
17 & DEEP2-3 & 23:31:25.2 & $+$00:35:20.2 & 4.86 & 28 & ELAIS-N1 & 16:13:50.2 & $+$55:44:37.5 & 4.53 \\
18 & ELAIS-N1 & 16:01:51.2 & $+$54:51:07.5 & 4.40 & 29 & ELAIS-N1 & 16:16:58.7 & $+$54:53:07.5 & 4.29 \\
19 & ELAIS-N1 & 16:02:43.6 & $+$54:37:37.5 & 4.07 & 30 & XMM-LSS & 02:17:07.5 & $-$04:22:25.9 & 4.53 \\
20 & ELAIS-N1 & 16:04:14.3 & $+$54:42:07.5 & 4.63 & 31 & XMM-LSS & 02:17:21.5 & $-$05:24:55.9 & 4.16 \\
21 & ELAIS-N1 & 16:04:14.3 & $+$54:51:37.5 & 4.19 & 32 & XMM-LSS & 02:18:17.7 & $-$05:28:55.9 & 4.08 \\
22 & ELAIS-N1 & 16:05:41.6 & $+$54:16:37.5 & 4.20 & 33 & XMM-LSS & 02:20:50.2 & $-$04:59:55.9 & 4.07 \\
23 & ELAIS-N1 & 16:06:09.5 & $+$55:58:37.5 & 4.00 & 34 & XMM-LSS & 02:21:04.3 & $-$04:00:25.9 & 5.03 \\
24 & ELAIS-N1 & 16:06:44.4 & $+$53:53:37.5 & 4.26 & 35 & XMM-LSS & 02:21:30.4 & $-$05:14:55.9 & 4.09 \\
25 & ELAIS-N1 & 16:06:58.4 & $+$55:01:37.5 & 4.13 & 36 & XMM-LSS & 02:22:00.5 & $-$03:42:25.9 & 4.12 \\
26 & ELAIS-N1 & 16:07:08.8 & $+$55:49:37.5 & 5.76 & 37 & XMM-LSS & 02:22:08.5 & $-$04:44:55.9 & 4.65 \\
27 & ELAIS-N1 & 16:08:29.1 & $+$53:37:37.5 & 5.40 & 38 & XMM-LSS & 02:23:14.7 & $-$05:13:55.9 & 4.49 \\
28 & ELAIS-N1 & 16:08:29.1 & $+$53:53:37.5 & 5.24 & 39 & XMM-LSS & 02:23:36.8 & $-$05:23:25.9 & 4.01 \\
29 & ELAIS-N1 & 16:12:54.4 & $+$56:40:07.5 & 4.12 & 40 & XMM-LSS & 02:24:06.9 & $-$05:36:25.9 & 4.11 \\
30 & ELAIS-N1 & 16:15:24.4 & $+$55:02:07.5 & 4.45 & 41 & XMM-LSS & 02:24:24.9 & $-$05:26:25.9 & 4.02 \\
31 & ELAIS-N1 & 16:16:13.3 & $+$54:59:37.5 & 4.81 & 42 & XMM-LSS & 02:26:15.3 & $-$04:16:55.9 & 4.08 \\
\hline
\end{tabular}
\end{table*}

\section{Model comparison} \label{sec:model}
To aid the interpretation of our observational analysis, we subject a theoretical light-cone model to the same
procedure of identification of protocluster candidates.
Since similar results are obtained for $U$-, $g$-, and $r$-dropout galaxies, implying at most modest protocluster
evolution from $z\sim5$ to $z\sim3$, we here present the results for the three epochs jointly.

In \citet{toshikawa16}, we investigated the relationship between surface overdensity of dropout galaxies and the
corresponding descendant halo mass at $z=0$ based on a set of light-cones constructed by \citet{henriques12}.
Since their light cones are constructed by tiling space with simulation boxes stored at discrete epochs,
particularly along the line-of-sight direction, galaxy properties are prone to discontinuously change at the
boundaries between contiguous boxes.
The updated light-cone model of \citet{henriques15} deals with this discontinuity by recording star-formation
histories at a finer time sampling.
Galaxy properties at any redshift can be extracted from this recorded star-formation history and galaxy spectra
similarly estimated by applying stellar population synthesis models in post-processing.
We refer the reader to \citet{shamshiri15} for further details on the construction of light cones based on the
recorded star-formation histories.
It should be noted that the latest light-cone model still does not perfectly reproduce all observational results,
especially at $z\gtrsim4$.
According to \citet{clay15}, the latest light-cone model provides a good agreement on SFR and stellar mass between
model predictions and observations even at high redshifts, but the predicted UV luminosity function of
high-redshift galaxies is a relatively poor fit to the observed luminosity function.
This deviation could be attributed to the incompleteness of stellar population synthesis and/or dust attenuation
models, although one should keep in mind that similar simplifying assumptions are also adopted in translating
direct observables to physical quantities such as SFRs and stellar masses when modelling the observations.
In this study, we update our model comparison to observed high-redshift overdense regions by using the latest
light-cone model of \citet{henriques15}.

The \citet{henriques15} model provides 24 different light cones, whose field-of-view (FoV) is a
$1\,\mathrm{deg}$-radius aperture, corresponding to a $75\,\mathrm{deg^2}$ area in total.
Various galaxy properties (e.g., halo mass, stellar mass, SFR) as well as their photometry in both rest and
observed frames have been made available via the public database of the Millennium simulation.
Due to the imperfect modelling of galaxy photometry or colours in the light cones, the mock catalogues of $U$-,
$g$-, and $r$-dropout galaxies are made by matching the expected redshift distribution of each dropout galaxy
sample rather than directly applying colour selection, as we did in \citet{toshikawa16}.
Then, the same overdensity measurement that was applied to the observational data is applied to the mock catalogues
of $U$-, $g$-, and $r$-dropout galaxies.
We introduce the fraction of protocluster members (progenitors of $>10^{14}\,\mathrm{M_{\sun}}$ halos) among all
dropout galaxies in an overdense region ($P_\mathrm{max}$) as the reliability of protocluster candidates.
If more than one protocluster are included in an overdense region due to chance alignments, $P_\mathrm{max}$ is
calculated only based on the protocluster which has the largest number of member galaxies.
In the simulated data, we found that 90.4\% of overdense regions contain at least one protocluster
($P_\mathrm{max}>0$), meaning that the rate of artificial protocluster candidates is less than 10\%.
However, in actual observations, it would be difficult to confirm the existence of protoclusters within overdense
regions if they had relatively small $P_\mathrm{max}$.
Therefore, observationally plausible candidates require higher $P_\mathrm{max}$ than a threshold value in order to
be reliably detected against the typical fluctuation in the number density of dropout galaxies not associated with
such structures.
Comparing the typical spread in redshift of protocluster members ($\Delta z\sim0.02\mathrm{-}0.04$) to the range
in redshift associated with a particular dropout selection
(typically $\Delta z\sim0.8\mathrm{-}1.0$ for $U$-, $g$-, or $r$-dropout galaxies), $P_\mathrm{max}=0.1$ would
result in a density of protocluster members that is at least twice higher than that of field dropout galaxies;
thus, overdense regions with $P_\mathrm{max}\gid0.1$ are plausible candidates.
On the other hand, it is difficult to observationally detect protoclusters from overdense regions with
$P_\mathrm{max}<0.05$ because such protoclusters are not composed of a sufficient number of member galaxies such
that spectroscopic follow-up of the overdense region would detect a clear spike in the redshift distribution
associated with the protocluster, despite its reality.
In summary, based on our model comparison, we have found that 76\% of protocluster candidates are plausible
($P_\mathrm{max}\gid0.1$) in terms of actual observations, 10\% still have a possibility of confirmation
($0.05\lid P_\mathrm{max}<0.1$), 4\% would not be confirmed observationally though they certainly include
galaxies that are members of a protocluster ($0<P_\mathrm{max}<0.05$), while 10\% are completely artificial
candidates ($P_\mathrm{max}=0$), just showing a surface overdensity of dropout galaxies which are spread over the
wide redshift range of $\Delta z\sim1$ and not associated with halos that are progenitors of a galaxy cluster at
$z=0$.
In contrast, when considering random locations on the sky (as opposed to $>4\sigma$ overdense regions), the
light-cone model suggest that only 11\% of such pointings would feature protocluster members with
$P_\mathrm{max}>0.1$.

We have also found that 16\% of overdense regions at high redshifts are dominated by the progenitors of galaxy
groups (here defined as $M_\mathrm{halo}=10^{13}\mathrm{-}10^{14}\,\mathrm{M_{\sun}}$ at $z=0$) rather than
protoclusters.
While these are straightforward to identify in simulations, they are a potential cause of contamination in the
observations.
Because of the stochastic nature of star and galaxy formation, it is possible that some high-redshift halos
destined to end up in groups at $z=0$ are richer in detectable galaxy members at an early epoch than other halos
destined to be part of clusters at $z=0$.
In addition, as noted by \citet{remus22}, the mass ranking of protoclusters at high redshifts is not necessarily
equal to that of descendant clusters at $z=0$, so progenitors of groups could be more massive than some of
protoclusters.
Furthermore, the observed number of protocluster members has some dependence on their redshifts because the
selection completeness of dropout galaxies varies within their selection window.
Nevertheless, although it is impossible to distinguish the rich progenitors of galaxy groups from protoclusters by
observations, the contribution of such contamination to our sample protocluster candidates is expected to be less
than 20\% as inferred from the mock light-cones.

Our sample of protocluster candidates was constructed with the aim of high purity despite the redshift windows of
dropout galaxies being much larger than the redshift size of protoclusters.
Unfortunately, optimising the selection for purity necessarily compromises the completeness of the sample.
Completeness is low because projection effects dilute the signal of galaxy clustering, leading to completeness
levels as assessed using the light-cones in the range of $6\mathrm{-}13\%$.
In particular, our observed sample of protocluster candidates is biased toward more massive protoclusters, with
the median descendant halo mass of our sample of protocluster candidates expected to be $\sim30\%\mathrm{-}50\%$
higher than that of a complete sample of protoclusters (including all of those systems that will end up in
clusters with masses down to $10^{14}\,\mathrm{M_{\sun}}$ at $z=0$).
From this model comparison based on the latest light-cone model by \citet{henriques15}, we obtained
consistent results on purity and completeness of our sample of protocluster candidates as those presented in
\citet{toshikawa16}, and we clarified the characteristics and possible caveats on the nature of our sample.

\section{Results} \label{sec:res}
Having described our data and protocluster search in Section~\ref{sec:data}, and the closely matched procedure
applied to mock light-cones in Section~\ref{sec:model}, we now turn to our observational results, which in
Section~\ref{sec:dis} will be discussed in the context of models and the observational literature.

\subsection{Number counts of dropout galaxies in overdense regions} \label{sec:numcount}
\begin{figure*}
\includegraphics[width=\textwidth,bb=0 0 1008 360]{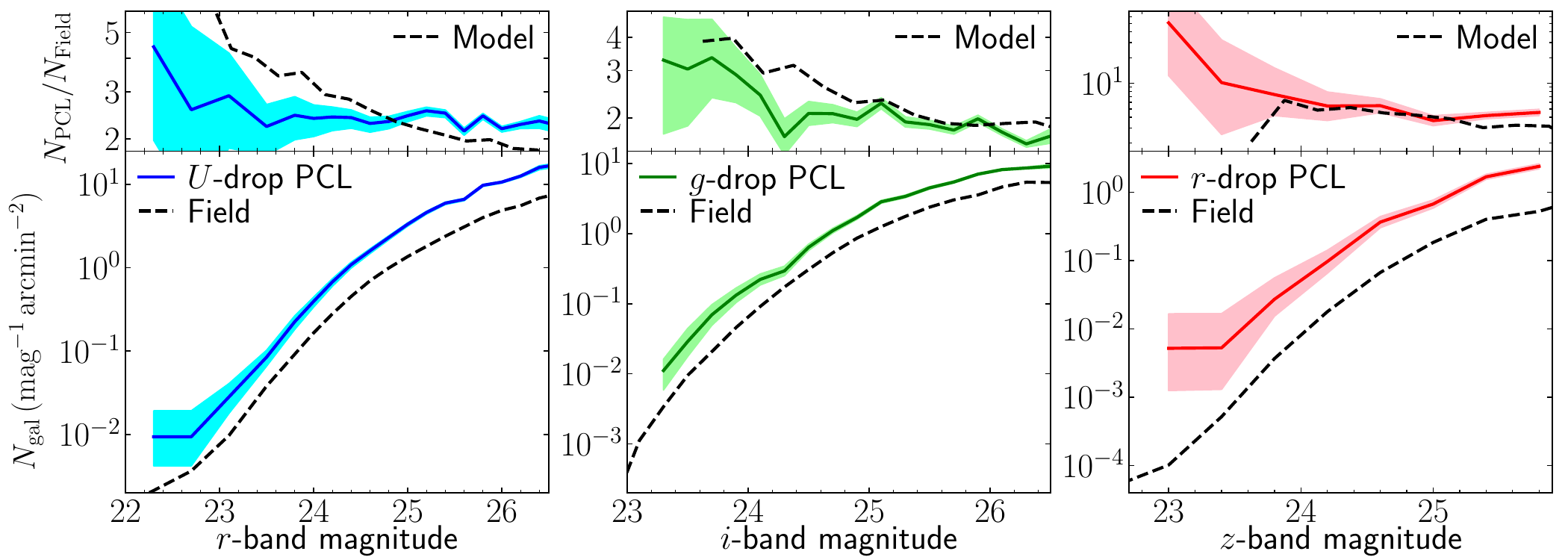}
\caption{Number counts of $U$-, $g$-, and $r$-dropout galaxies from left to right panels.
    The coloured and black lines are for overdense and field regions, respectively.
    The $1\sigma$ uncertainties of number counts in overdense regions are indicated by shaded regions.
    The smaller upper panels show the ratio of number counts between overdense and field regions as a function of
    apparent UV magnitude.
    The dashed lines in the upper panels are the model predictions.}
\label{fig:numcount}
\end{figure*}

First, we start by comparing the number counts of dropout galaxies in the regions of protocluster candidates with
those in the remaining area of the DUD layer.
Our aim is to probe possible differences between protocluster and field galaxies at $z\sim3\mathrm{-}5$.
The fact that both samples are constructed from the same dataset represents a strength of our analysis, as any
effects originating from observational and/or selection biases would apply equally to both samples of field and
protocluster candidates, and are therefore expected to impact the comparison negligibly.
Figure~\ref{fig:numcount} compares the number counts of dropout galaxies in the regions ($>2\sigma$ overdensity)
of protocluster candidates and the average field environment.
In this study, we focus on the shape of the number count distribution, instead of directly
comparing amplitudes between protocluster and field galaxies due to the following uncertainties or systematic
effects on the estimate of galaxy number density.
Firstly, the factor by which the surface number densities of overdense regions exceed that of the field depends
on survey depth.
That is, the depth of observations determines both the average number of field galaxies within our
$0.75\,\mathrm{pMpc}$ sized apertures and its standard deviation, and thus affects which regions satisfy the
adopted definition of a protocluster candidate ($>4\sigma$ overdensity).
In addition, surface overdensity can be increased by the chance alignment of small groups along the line-of-sight
direction.
The average surface number density in overdense regions is $>2$ times higher than that in field environments (the
upper panels of Figure~\ref{fig:numcount}).
Although this means that more than half of all galaxies in overdense regions cause the excess of surface number
density, only $\sim20\%$ of them are expected to be the member galaxies of a single protocluster.
Furthermore, it is difficult to estimate the volume enclosing protocluster members.
In the early phase of cluster formation, the spatial distribution of protocluster members tend to be filamentary
or could be divided into some fragments \citep[e.g.,][]{lovell18,toshikawa20}.
Thus, the amplitude of surface number density in overdense regions is related to the dataset and the definition of
protocluster candidates, and the estimate of their volume density involves many assumptions.

The number counts of overdense regions normalised by that of field galaxies are presented in the upper panels of
Figure~\ref{fig:numcount}.
In $g$- and $r$-dropout galaxies, overdense regions exhibit a clear upturn of their excess number density at the
bright end.
At the fainter end, on the other hand, the ratio of number counts remains almost constant.
A similar trend is also found in $U$-dropout galaxies though their larger uncertainties make the upturn less
significant.
The left panel of Figure~\ref{fig:ratios} combines the ratios of number counts for all redshift samples as a
function of absolute UV magnitude, $M_\mathrm{UV}$.
The magnitude where the excess starts to rise is seen around $M_\mathrm{UV}\sim-21.5\,\mathrm{mag}$.
These results suggest that brighter galaxies are relatively more abundant (by a factor of $\sim2$) in protocluster
candidates at $z\sim3$ to $z\sim5$.
A qualitatively similar bright-end excess was found at $z\sim4$ in the HSC-Wide layer \citep{ito19}, and this
paper has confirmed that this trend is common from $z\sim3$ to $z\sim5$.

\begin{figure*}
\includegraphics[width=\textwidth,bb=0 0 1080 288]{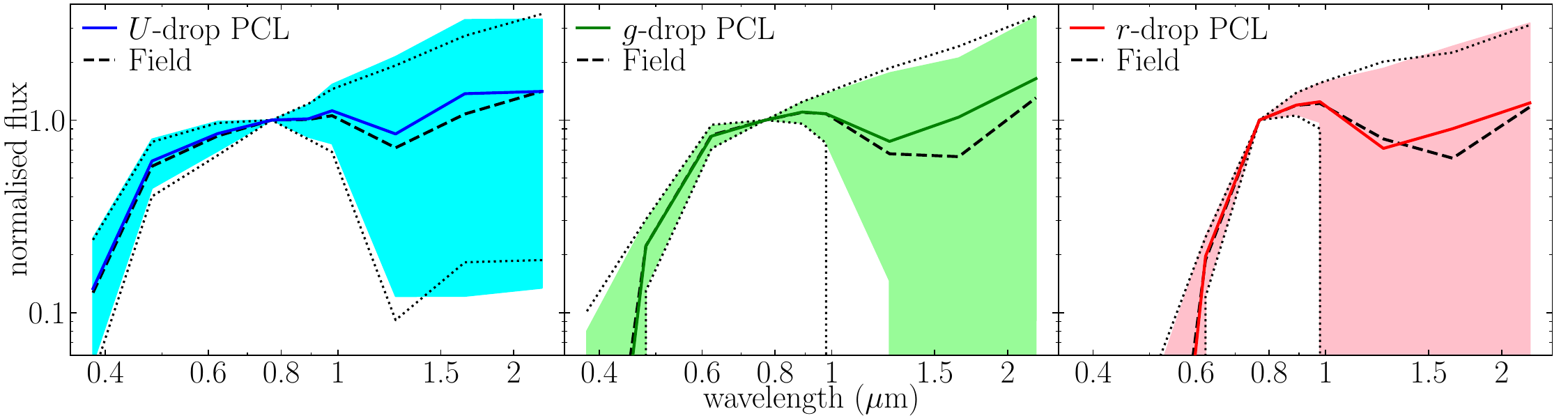}
\caption{Median SEDs of $U$-, $g$-, and $r$-dropout galaxies from left to right panels.
    Fluxes are normalised by the $i$-band flux.
    The coloured and black lines are for objects in overdense and field regions, respectively.
    The area between upper and lower quartiles in overdense regions are indicated by shaded polygons, and the upper
    and lower quartiles in the field are presented by the dotted lines.}
\label{fig:aveSED}
\end{figure*}

\begin{figure*}
\includegraphics[scale=0.46,bb=0 0 1100 340]{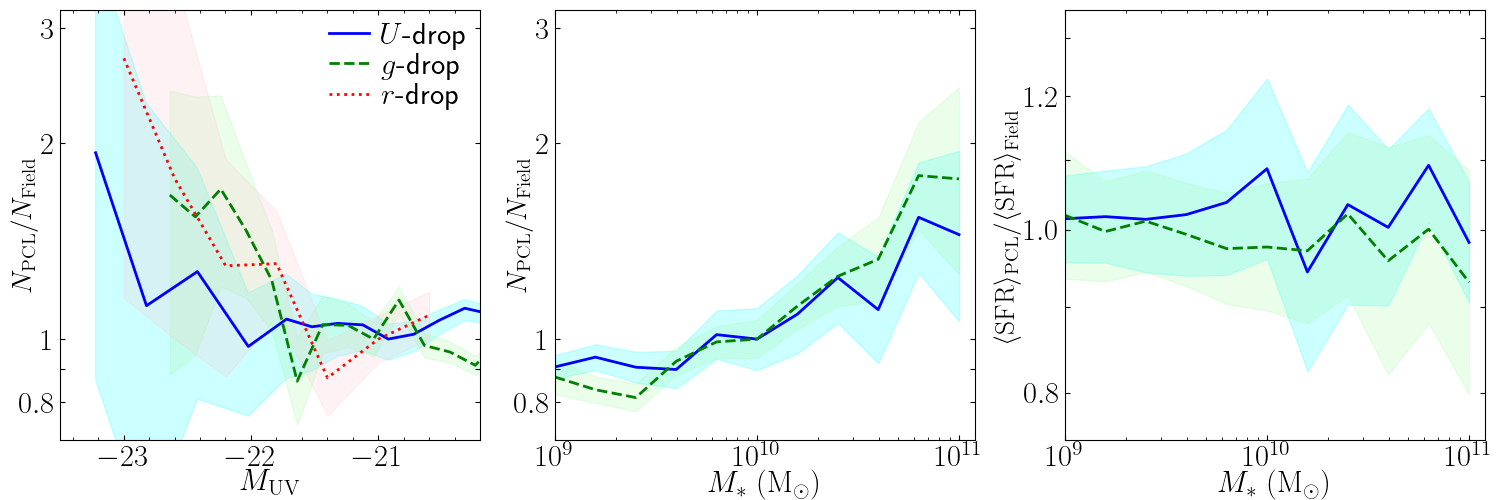}
\caption{{\it Left:} the ratio of number counts between overdense and field regions as a function of absolute UV
    magnitude.
    The blue, green, and red lines are $U$-, $g$-, and $r$-dropout galaxies, respectively.
    Although the ratios for all redshift samples are normalised at $M_\mathrm{UV}=-21.0\,\mathrm{mag}$, the graphs
    are the same as in the upper panels in Figure~\ref{fig:numcount}.  Polygons denoting uncertainties are based
    on Poisson statistics.
    {\it Middle:} the same as in the left panel, but as a function of stellar mass derived by SED fitting.
    The ratios for both $U$- and $g$-dropout galaxies are normalised at $M_{\ast}=10^{10}\,\mathrm{M_{\sun}}$.
    {\it Right:} the ratio of average SFR between overdense and field regions as a function of stellar mass.
    The uncertainties indicated by polygons are derived by bootstrapping and incorporate both sample variance and
    errors from SED fitting.}
\label{fig:ratios}
\end{figure*}

We also investigate the relative number counts of dropout galaxies in overdense and field regions based on the
light-cone model.
These results are depicted by the dashed lines in the upper panels of Figure~\ref{fig:numcount}.
The model predicts an enhanced excess at the bright end in $U$- and $g$-dropout galaxies, but features a paucity of
bright galaxies in $r$-dropout overdense regions.
Although the upturn of the excess of number density is qualitatively reproduced by the light-cone model, there are
some inconsistencies between observation and model from a quantitative viewpoint.
In this context, we remind the reader that the model analysed here does not yet completely reproduce field galaxies
themselves either (see Section~\ref{sec:model}).
Previous comparisons between models and observations of protoclusters or high-density environments have similarly
reported (sometimes significant) quantitative differences \citep[e.g.,][]{lim21,remus22}.

In actual observations, objects at close angular separation may be confused into a single source due to seeing
convolution.
Using the light-cone model, we evaluated that this confusion effect is only able to increase galaxy numbers at the
bright end by a few percent at maximum, even in overdense regions.
Furthermore, a systematic excess over a range of $\sim1\,\mathrm{mag}$ as seen in Figure~\ref{fig:numcount} cannot
be reproduced by the random phenomenon of source confusion.
Given the results from our sanity check on the mock light-cone, the effect of source confusion on galaxy number
counts is not considered in our observational analysis.

\subsection{Galaxy properties derived by SED fitting} \label{sec:sed}
We now investigate the physical properties of dropout galaxies in overdense regions based on multi-wavelength
$U$-to-$K(K_s)$ data.
We do this with the aim of understanding the cause of the upturn at the bright end.
Figure~\ref{fig:aveSED} shows the median normalised SEDs for $U$-, $g$-, and $r$-dropout galaxies in overdense
regions, where the $i$ band was chosen as normalisation point, for reasons of its superior depth.
No clear difference in colours is discernible between galaxies in overdense regions and the field; their median
SED shapes being determined largely by the definition, or colour criteria, of dropout selection (i.e., a strong
Lyman break and flat UV continuum).
We perform further detailed investigations by using the SED fitting code CIGALE
\citep{burgarella05,noll09,boquien19}.
As the wavelength coverage of our dataset is not wide enough to estimate the physical properties of $r$-dropout
galaxies, we apply SED fitting to $U$- and $g$-dropout galaxies only.
CIGALE offers great flexibility to model the SEDs of galaxies.
The set of broad-band fluxes of various galaxy models are compared to the observed fluxes to determine a
best-fitting SED model and the corresponding physical quantities (e.g., stellar mass, SFR, or age) by $\chi^2$
analysis.
In this study, instead of adopting the physical parameters corresponding to the model with the minimum $\chi^2$,
a likelihood-weighted mean and standard deviation are used as the fitting results of value and error, where
likelihood is calculated as $\exp(-\chi^2/2)$.

We briefly describe which models/modules of CIGALE are employed in this study.
We have used a delayed star-formation history (SFH), which is expressed by
$\mathrm{SFR}(t) \propto t/\tau^2 \times \mathrm{exp}(-t/\tau)$ with $\tau$ representing the time at which the SFR
peaks, and $t$ the age since the onset of star formation.
The stellar emission is computed using the population synthesis model of \citet{bruzual03} with a
\citet{salpeter55} initial mass function (IMF).
Stellar metallicity is allowed to be either $0.2\,Z_{\sun}$ or $0.4\,Z_{\sun}$, and the dust attenuation law
adopted is \citet{calzetti00}.
Finally, model SEDs are redshifted accounting for IGM absorption following the prescription by \citet{meiksin06}.
Since our multi-wavelength dataset consists of only broad-band imaging at up to rest-optical wavelengths, the
degeneracy between SFH and dust attenuation cannot be fully resolved with such SED fitting approach.
Thus, we need to fix or narrow parameter ranges in order to avoid unrealistic best-fitting SED models.
We set $\tau=10\,\mathrm{Gyr}$ in order to force dropout galaxies into a constant or slowly rising SFH, while
the age of dropout galaxies can range from $25\,\mathrm{Myr}$ to the age of the universe at the observed redshift.
The parameter range of dust reddening is $E(B-V)=0.0\mathrm{-}0.3$, and redshift is left to vary freely between
$2.7\mathrm{-}3.7$ for $U$-dropout and $3.3\mathrm{-}4.3$ for $g$-dropout galaxies, respectively, corresponding to
the redshift windows of each dropout sample.

The completeness of stellar mass can be empirically estimated as follows \citep[e.g.,][]{pozzetti07,laigle16}:
$\log(M_\mathrm{\ast,lim})=\log(M_{\ast})-0.4(K_\mathrm{lim}-K)$, where $M_\mathrm{\ast,lim}$ is the stellar mass
limit, $M_{\ast}$ is the derived stellar mass, $K_\mathrm{lim}$ is the $K$-band limiting magnitude, and $K$ is the
observed $K$-band magnitude.
The 90\% completeness of stellar mass is estimated to be $\sim7.0\times10^9\,\mathrm{M_{\sun}}$ and
$\sim1.2\times10^{10}\,\mathrm{M_{\sun}}$ for $U$- and $g$-dropout galaxies, respectively, in the shallowest
regions ($K_\mathrm{lim}\sim23.0\,\mathrm{mag}$).
The typical errors for stellar mass and SFR are $\sim30\%$ and $\sim60\%$, respectively.

The middle and right panels of Figure~\ref{fig:ratios} show the ratio of number counts as a function of stellar
mass and the ratio of average SFR at a given stellar mass between overdense regions and field, respectively.
The ratios of number counts are normalised at $M_{\ast}=10^{10}\,\mathrm{M_{\sun}}$.
In both overdense regions of $U$- and $g$-dropout galaxies, massive galaxies are found to be more abundant relative
to the field, resembling the bright-end excess shown in Section~\ref{sec:numcount} or the left panel of
Figure~\ref{fig:ratios}.
The amplitude of the excess is also comparable.
On the other hand, there is no significant difference in terms of SFR at fixed mass between overdense and field
regions.
These results suggest that the rich population of UV-bright galaxies is caused by a higher abundance of massive
galaxies rather than a more enhanced SFR (at fixed mass) in overdense regions.
It should be noted that, based on our multi-wavelength dataset, SFR cannot be precisely constrained and SED fitting
is unreliable for $r$-dropout galaxies.
In Section~\ref{sec:dis}, we will further investigate environmental trends in star formation and galaxy stellar
mass distributions over the full redshift range, by making use of theoretical models.

\begin{figure*}
\includegraphics[width=\textwidth,bb=0 0 1008 360]{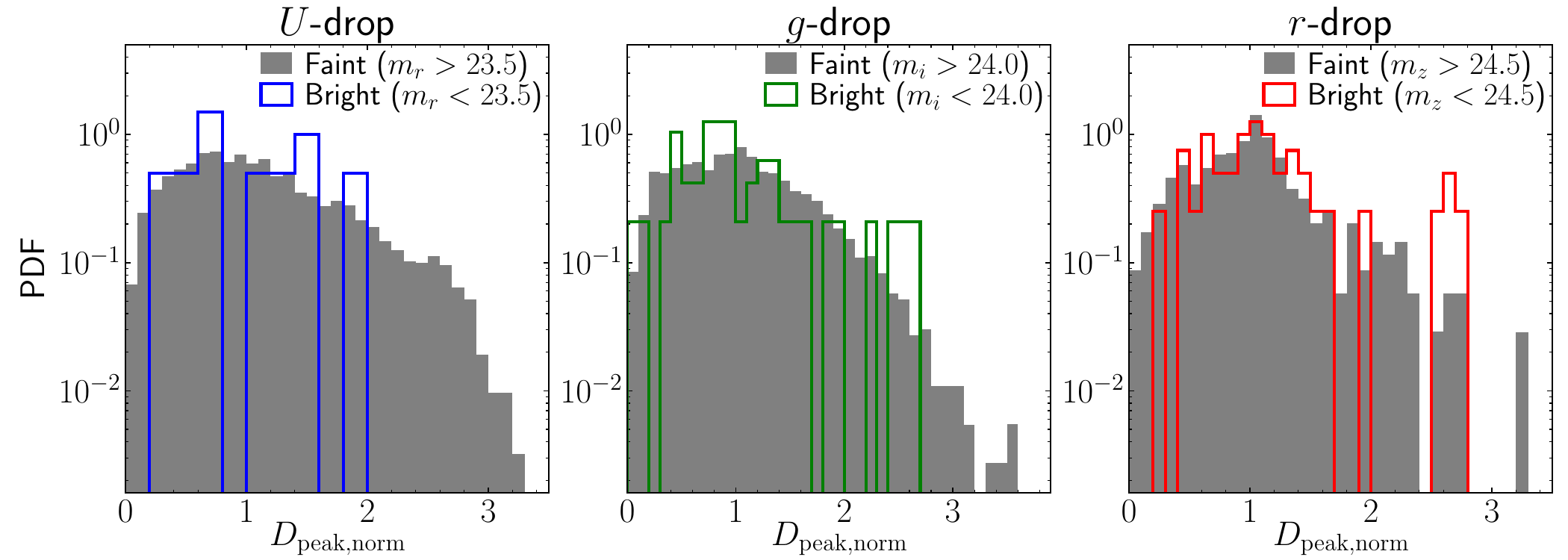}
\caption{Radial distribution of $U$-, $g$-, and $r$-dropout galaxies in overdense regions from left to right
    panels.
    The distributions of bright and faint galaxies are indicated by coloured solid lines and the underlying gray
    histogram, respectively.
    The radial distance from overdense peaks is normalised by the median distance of galaxies in each overdense
    region.}
\label{fig:Dcen}
\end{figure*}

\subsection{Locations of bright galaxies in overdense regions} \label{sec:pos-brigal}
We now investigate the spatial location of bright galaxies within overdense regions, focussing on the magnitude
range where the higher excess in surface number density is observed ($m_r<23.5\,\mathrm{mag}$,
$m_i<24.0\,\mathrm{mag}$, and $m_z<24.5\,\mathrm{mag}$ for $U$-, $g$-, and $r$-dropout galaxies, respectively).
To this end, we measure for each dropout galaxy residing within a $>2\sigma$ overdense region around a $>4\sigma$
overdensity peak the distance to this peak.
To account for the fact that the spatial extent of overdense regions varies, we normalise for each protocluster
candidate the individual distance measurements by the median distance to the overdensity peak.
Figure~\ref{fig:Dcen} shows the probability distribution function (PDF) of such normalised distances from
overdensity peaks, $D_\mathrm{peak,\ norm}$.
There is no difference in the $D_\mathrm{peak,\ norm}$ PDFs between bright and faint galaxies in overdense regions
according to the Kolmogorov-Smirnov test.
Alternative ways of normalising (or not) were explored, but did not reveal a spatial differentiation between
bright and faint galaxies within overdense regions either.
Neither did a quantification of spatial distribution based on distances to the nearest neighbour, an approach that
may have been more appropriate if protoclusters were composed of several independent halos with bright galaxies
residing at the centres of such substructures.
The lack of a significant spatial differentiation owes in part to the large sample variation, and at face value
implies that at these early stages of development, protoclusters have not yet formed centrally concentrated
structures.
In most cases, dark matter halos which will merge into single massive halos like galaxy clusters are still spread
over a wide area ($\sim10\,\mathrm{comoving\>Mpc}$) at $z\sim3\mathrm{-}5$ \citep[e.g.,][]{muldrew15,chiang17}.

Our results suggest that the spatial distribution of bright galaxies is not significantly biased in protocluster
candidates at $z\ga3$, although overdense regions do show a higher excess of galaxy number counts at the brighter
end.
It should be noted that contamination by foreground and background galaxies can influence the above analysis.
Due to projection effects, the peak position of surface overdensity can be shifted by as much as
$\sim1\,\mathrm{arcmin}$ from the centre of a protocluster based on the light-cone model.
Given the typical $P_\mathrm{max}$ of $\sim0.1\mathrm{-}0.2$ inferred from the light-cone model, it is likely
that nearest neighbour dropout galaxies are not physically associated with each other.
Furthermore, in the early stage of cluster formation, it is possible that a strong correlation between galaxy
properties and environments has not yet emerged \citep[e.g.,][]{malavasi21,lemaux22}.
More subtle levels of spatial differentiation would need a larger number of protocluster candidates to be
detected, or -ideally- could be revealed via investigations of their 3D distributions as enabled by spectroscopic
follow-up \citep{cucciati18,toshikawa20}.

\section{Discussion} \label{sec:dis}
\subsection{Physical origin of bright-end excess}
We have found that the ratio of dropout galaxy number counts between overdense and field regions increases at the
bright end.
It suggests that environmental effects work preferentially on or are preferentially expressed in brighter galaxies.
Physically, the observed UV luminosities and variations therein can in principle stem from the galaxies'
star-formation and/or AGN activity, and may further be modulated by dust attenuation.
The UV-continuum slope is a good indicator of the level of dust extinction \citep[e.g.,][]{bouwens12}.
To quantify this parameter, we make use of the $r-i$ and $i-z$ colour for $U$- and $g$-dropout galaxies,
respectively.
Due to the shallower depth of $y$-band imaging in the HSC-SSP, we adopt the $i-z$ colour instead of the $z-y$
colour to determine the UV colour of $r$-dropout galaxies.
Although the $i-z$ colour of $r$-dropout galaxies is additionally affected by IGM absorption and potentially by
strong Ly$\alpha$ emission, if present, it is still worth comparing $i-z$ colours between overdense and field
regions in order to identify possible differences in dust extinction.
We find the UV colour distribution of galaxies in overdense regions to be consistent with that of coeval field
galaxies for all redshift samples.
This indicates that dust extinction does not cause the bright-end excess in overdense regions.

AGN contribution to the bright-end excess is also expected to be negligible or minor because the magnitude range
where the number count excess is increasing ranges from $M_\mathrm{UV}\sim-23\,\mathrm{mag}$ to
$M_\mathrm{UV}\sim-22\,\mathrm{mag}$.
According to \citet{harikane22}, this magnitude range is still dominated by galaxies' stellar emission rather than
AGN.
Actually, only one AGN is found from six spectroscopically-confirmed protoclusters at $z\sim3\mathrm{-}5$
\citep{toshikawa16,toshikawa20}, which are selected via an identical method as applied in this study.
The $i$-band magnitude of that AGN at $z=3.7$ is $23.2\,\mathrm{mag}$, which is brighter than the magnitude range
of the bright-end excess reported in this paper.
It should be noted that a larger number of faint AGNs in protoclusters remains possible.
Deep X-ray or spectroscopic observations will be required to assess this possibility.

\begin{figure}
\includegraphics[width=\columnwidth,bb=0 0 432 432]{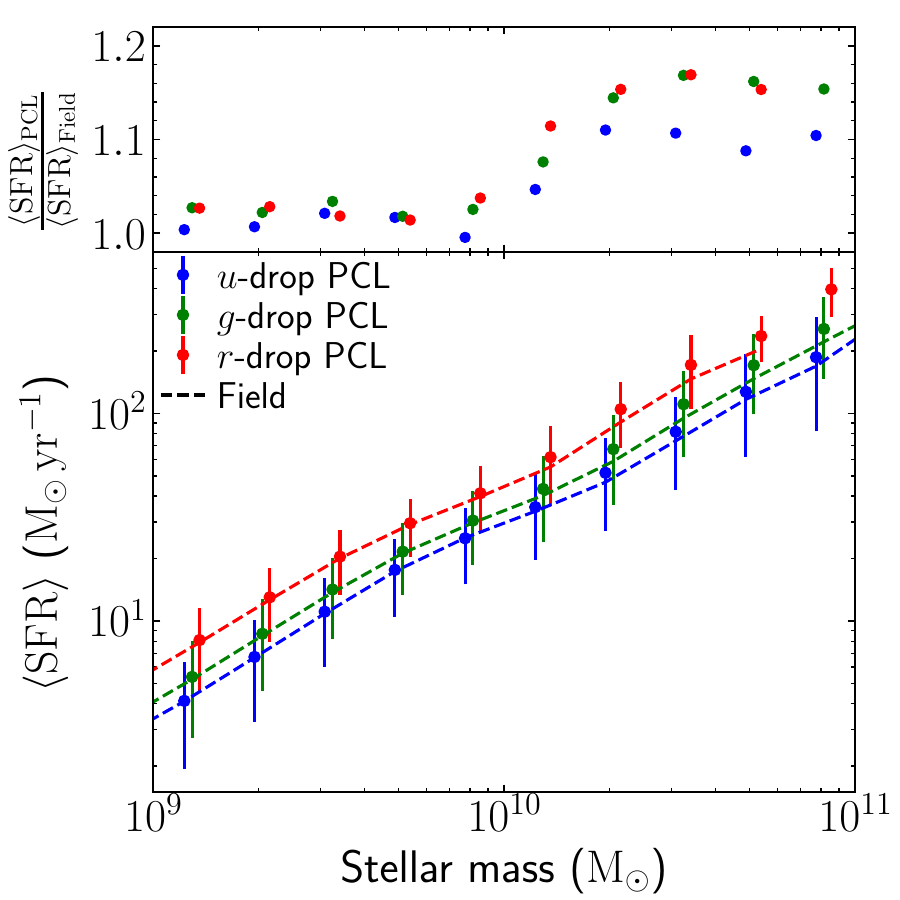}
\caption{Model predictions of median SFR of protocluster (points with error bars) and field
    galaxies (dashed lines) as a function of stellar mass (blue: $U$-dropout, green: $g$-dropout, and red:
    $r$-dropout galaxies).
    The upper panel shows the SFR ratio between protocluster and field galaxies.}
\label{fig:modelMS}
\end{figure}

Finally, protocluster galaxies may have a different star-formation activity from field
galaxies.
The results obtained by SED fitting (Section~\ref{sec:sed}) imply that there is no difference in star-formation
activity at fixed stellar mass, but that the galaxy population in protocluster candidates is biased toward more
massive galaxies.
We consider this scenario further in the context of the theoretical model of \citet{henriques15}.
In Figure~\ref{fig:modelMS}, the model SFRs of protocluster galaxies are compared with that of field galaxies in
order to seek a possible cause of the enhanced bright-end excess.
The model features no difference in SFR at a given stellar mass between protocluster and field galaxies in the low
stellar-mass range of $M_\ast\la 10^{10}\,\mathrm{M_{\sun}}$, while protocluster galaxies form
$\sim10\%\mathrm{-}20\%$ more stars than equal-mass counterparts in the field at the higher stellar-mass end.
Note that this modest yet systematic offset has an amplitude smaller than the scatter in specific SFRs (SFR/$M_*$)
among either protocluster or field populations.
As SFR is proportional to UV luminosity \citep[e.g.,][]{kennicutt98}, this SFR enhancement contributes to making
their UV magnitudes modestly brighter.
Interestingly, the ratio of median SFR between protocluster and field galaxies at fixed stellar mass jumps at
$\sim1\times10^{10}\,\mathrm{M_{\sun}}$, which appears to be consistent with the upturn of the ratio of number
density between protocluster and field galaxies as a function of UV magnitude (the upper panels of
Figure~\ref{fig:numcount}).

\begin{figure}
\includegraphics[width=\columnwidth,bb=0 0 432 360]{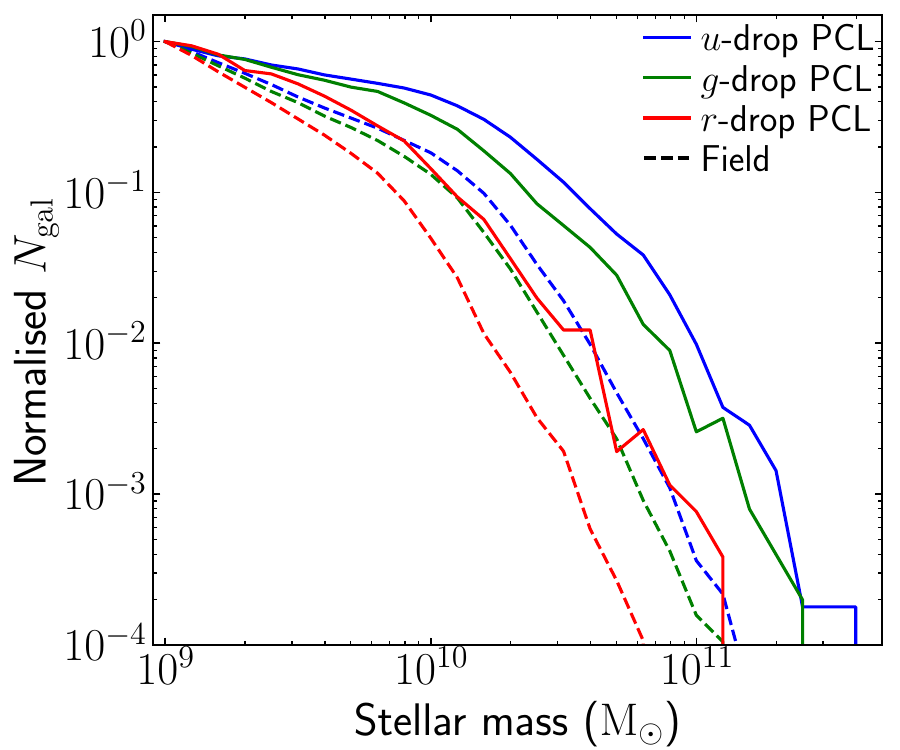}
\caption{Model predictions of stellar mass functions of $U$-, $g$-, and $r$-dropout galaxies, normalised at
    $10^9\ M_{\odot}$.
    The solid and dashed lines indicate that of protocluster and field galaxies, respectively.}
\label{fig:modelMstar}
\end{figure}

More importantly, star-forming galaxies populate a tight relation between stellar mass and SFR, the so-called main
sequence; thus, the bright-end excess may also be due to a richer massive galaxy population with consequently
higher SFRs, even if following a similar main sequence relation.
We have examined the model prediction for how the shape of the galaxy stellar-mass function depends on environment.
Massive galaxies are more abundant in protoclusters (Figure~\ref{fig:modelMstar}).
The model shows a gradual increase in the ratio of protocluster over field galaxies starting already below
$M_\ast\la10^{10}\,\mathrm{M_{\sun}}$, and becoming more dramatic above the knee of the mass function.
Phrased another way, the stellar-mass function of $r$-dropout protocluster galaxies shows a comparable shape with
that of lower-redshift field samples of $U$- or $g$-dropout galaxies, implying an accelerated galaxy growth in
high-density environments.

Thus, the theoretical model predicts SFRs that at fixed mass are up by only $10\mathrm{-}20\%$ and only so above
$10^{10}\,\mathrm{M_{\sun}}$.
At the same time, a significantly different shape of the galaxy stellar-mass function is revealed, over a wide
range in masses.
This points at galaxy mergers speeding up galaxy evolution in high-density environments, while leaving the
individual galaxy properties nearly identical to those of equal-mass field counterparts, at least among
UV-selected protocluster candidates at $z\gtrsim3$.
The fact that observationally no difference between the colours of galaxies in overdense and field regions was
revealed may mean that galaxy mergers, happening more frequently in higher density environments, largely leave the
underlying stellar populations unchanged, but simply increase the amount of stellar mass.
Based on an analysis of shapes of quiescent galaxies at $0<z<0.9$, \citet{zhang22} concluded that massive
elliptical galaxies in high-density regions are built up through more frequent mergers.
\citet{sawicki20} found that ultra-massive passive galaxies at $z\sim1.6$ are accompanied by a
smaller number of satellites, suggesting that the progenitors of brightest cluster galaxies are evolving at a
higher rate in their earlier phase, after which their growth rate decreases, having consumed their massive
companions already.
We thus conclude that both SED modelling of observed galaxies and analysis of the simulated galaxy population
attribute a greater role to a larger abundance of massive galaxies than a larger abundance of starbursting (high
${\rm SFR}/M_*$) systems in explaining the bright-end excess of UV number counts.
The cosmic clock appears to be ahead in high-density environments, plausibly due to earlier galaxy formation and
more frequent mergers.

\subsection{Comparison with other protoclusters}
Interpreted jointly with model predictions, our results do not immediately suggest a (strong) enhancement of
star-formation activity (in terms of SFR/$M_*$) in high-density environments.
Instead, the bright-end excess found in protocluster candidates is more naturally explained in terms of a more
richly populated massive end of the main sequence of star-forming galaxies.
I.e., they correspond to regions where the cosmic clock is ahead.
The same trend was found for some protoclusters at $z\sim2$ \citep{hatch11,koyama13}, while a protocluster which
is dominated by quiescent galaxies was found even at $z=3.37$ \citep{mcconachie22}.
However, there have also been discoveries of protoclusters including many sub-mm galaxies (SMGs), or dusty
star-bursts \citep{miller18,oteo18}, indicating that star-formation activity is strongly enhanced in
protoclusters.
Although this variety of star-formation activity in protoclusters may reflect the inherent diversity of physical
properties of protoclusters, it is also possible that it can be (partly) attributed to the specific observational
biases of different protocluster searches. 
For example, phot-$z$ selection inherently has a sensitivity to quiescent galaxies, with well-pronounced breaks
in their SED, while young star-forming galaxies tend to be more featureless in their SEDs except for emission
lines (if probed spectroscopically or via narrow-band selection) and their Lyman break (if not hidden by dust
attenuation).
On the other hand, the SMG-dominated protoclusters were initially identified as bright sub-mm sources in the
extremely wide-area surveys by Herschel and the South Pole Telescope (SPT).
Such protoclusters may be the rare and unique examples of protoclusters that break the general rule, as their
galaxy populations do appear distinct from equal-mass galaxies in the field.
It should be noted that \citet{rotermund21} performed follow-up optical imaging for a protocluster composed of
many SMGs and found only a marginal overdensity in terms of the number density of $g$-dropout galaxies.
Furthermore, they found $g$-dropout galaxies with redder UV slopes to be more concentrated near the protocluster
core.
As the dropout technique is more sensitive to UV-bright or dust-poor galaxies, this study may miss protoclusters
which are predominantly filled with dusty galaxies.
Alternatively, the dusty starburst phase may represent a short-lived stage in protocluster evolution.
Therefore, it is difficult to construct a general picture of protoclusters from comparisons to select
serendipitous examples.
Our results based on a systematic sample suggest that protoclusters have a consistent main-sequence of
star-forming galaxies with that of field galaxies but host a galaxy population that is biased toward the massive
end.

This protocluster candidate study is based on a dropout-selected population of galaxies within surface overdense
regions on the sky.
The redshift uncertainty of the dropout selection technique (projection of $\Delta z\sim1$) implies that recovered
trends will be diluted compared to those present in real 3D space.
Without projection effects, \citet{lemaux22} carried out a detailed investigation of the relation between SFR and
environments at $2<z<5$ by using spectroscopy from the VIMOS Ultra-Deep Survey \citep[VUDS:][]{lefevre15}.
They found that galaxies in higher-density environments exhibit a higher SFR.
Similar to our findings, this positive correlation between SFR and environments is primarily accounted for by the
increase of stellar mass in protocluster galaxies, although even when controlling for mass these authors find a
weak but significant SFR -- overdensity trend to remain present.
Although the spectroscopic survey enabled a precise measurement of galaxy density without contaminating effects
from foreground and background galaxies, the number of high-density environments, or protoclusters, identified
within their study remained limited to a few in total.
This systematic spectroscopic survey also found accelerated stellar-mass evolution and comparatively weak
enhancements in SFR at a given mass in higher-density environments.
Our complementary study based on wide-field imaging constructs a large sample of protocluster candidates, allowing
us to investigate protocluster properties per unit of redshift from $z\sim5$ to $z\sim3$.
We have confirmed a bright-end excess even at $z\sim5$, pushing the onset of protoclusters' accelerated growth
to yet earlier cosmic epochs.

\section{Conclusions} \label{sec:con}
We have presented a systematic search for protoclusters across cosmic time ($z\sim3\mathrm{-}5$) in the Deep and
UltraDeep layer of the HSC-SSP, resulting in a sample of over 100 protocluster candidates.
The same procedures are applied to light-cone models in order to interpret the physical properties of protoclusters
imprinted in observational results.
The results and implications of this study are summarised as follows:

\begin{enumerate}
\item At the bright-end ($M_\mathrm{UV}\la-21.5$), galaxies are more abundantly populating overdense regions than
    field environments, while the ratio of galaxy number between overdense and field regions stays flat at fainter
    UV magnitudes.
    This bright-end excess is common in all three redshift samples.
\item Multi-wavelength SED fitting reveals no significant difference between physical properties of dropout
    galaxies in overdense and field regions when compared at fixed stellar mass.
    However, the galaxy population in overdense regions is skewed to higher masses, naturally explaining the
    observed bright-end excess in number counts.  
    An equivalent analysis on mock light-cones shows that this is qualitatively consistent with model predictions.
    Galaxies residing in high-density environments follow a similar SFR -- stellar mass relation as coeval field
    populations.
    Although the star-forming activity of protocluster galaxies in the model is enhanced at the massive end, the
    absolute increase of SFR at fixed mass is small ($\sim15\%$).
    Therefore, the bright-end excess is caused mainly by a larger abundance of massive galaxies,
    implying an accelerated evolution of protocluster galaxies compared to the average field environment.
\item Despite their unique feature in the number-count or stellar-mass distributions, such bright galaxies are
    randomly distributed over overdense regions.
    This likely reflects that protoclusters are dynamically unrelaxed and member galaxies reside in independent
    halos.
    After all, our protocluster search based on high-redshift dropout galaxies traces the early phase of cluster
    formation.
\end{enumerate}

Since this study is based on imaging data without spectroscopic redshifts, contamination by foreground and
background galaxies makes it challenging to quantitatively evaluate the differences of physical properties between
protocluster and field galaxies.
However, our results confirm that environmental differences of galaxy properties appear even at $z\sim5$,
stressing the importance of tracing the history of cluster formation for understanding galaxy evolution,
and especially the formation of massive galaxies.
Unlike previous protocluster studies that focussed on a few protoclusters, this conclusion is drawn from a large
systematic sample.
In the future, we will carry out further follow-up observations to unveil the physical mechanisms which are
unique in high-density environments.

\section*{Acknowledgements}
We thank the anonymous referee for valuable comments and suggestions that improved the manuscript.
JT and SW acknowledge support from STFC through grant ST/T000449/1.
CL acknowledges support from the National Natural Science Foundation of China (NSFC, Grant No.\,12173025,
11833005, 11933003), 111 project (No.\,B20019), and Key Laboratory for Particle Physics, Astrophysics and
Cosmology, Ministry of Education.
MK acknowledges supports from JSPS KAKENHI Grant Numbers  20K14530 \& 21H044902 and Tohoku University Center for
Gender Equality Promotion (TUMUG).
TK acknowledges financial support from JSPS KAKENHI Grant Number 18H03717.
The HSC collaboration includes the astronomical communities of Japan and Taiwan, and Princeton University.
The HSC instrumentation and software were developed by the National Astronomical Observatory of Japan (NAOJ), the
Kavli Institute for the Physics and Mathematics of the Universe (Kavli IPMU), the University of Tokyo, the High
Energy Accelerator Research Organization (KEK), the Academia Sinica Institute for Astronomy and Astrophysics in
Taiwan (ASIAA), and Princeton University.
Funding was contributed by the FIRST program from Japanese Cabinet Office, the Ministry of Education, Culture,
Sports, Science and Technology (MEXT), the Japan Society for the Promotion of Science (JSPS), Japan Science and
Technology Agency (JST), the Toray Science Foundation, NAOJ, Kavli IPMU, KEK, ASIAA, and Princeton University.
This paper makes use of software developed for the Large Synoptic Survey Telescope (LSST).
We thank the LSST Project for making their code available as free software at \url{http://dm.lsst.org}.
These data were obtained and processed as part of the CFHT Large Area $U$-band Deep Survey (CLAUDS), which is a
collaboration between astronomers from Canada, France, and China described in \citet{sawicki19}.
CLAUDS is based on observations obtained with MegaPrime/MegaCam, a joint project of CFHT and CEA/DAPNIA, at the
CFHT which is operated by the National Research Council (NRC) of Canada, the Institut National des Science de
l'Univers of the Centre National de la Recherche Scientifique (CNRS) of France, and the University of Hawaii.
CLAUDS uses data obtained in part through the Telescope Access Program (TAP), which has been funded by the
National Astronomical Observatories, Chinese Academy of Sciences, and the Special Fund for Astronomy from the
Ministry of Finance of China.
CLAUDS uses data products from TERAPIX and the Canadian Astronomy Data Centre (CADC) and was carried out using
resources from Compute Canada and Canadian Advanced Network For Astrophysical Research (CANFAR).

\section*{Data Availability}
The data underlying this article will be shared on reasonable request to the corresponding author.

\bibliographystyle{mnras}
\bibliography{refs}

\begin{thebibliography}{}
\makeatletter
\relax
\def\mn@urlcharsother{\let\do\@makeother \do\$\do\&\do\#\do\^\do\_\do\%\do\~}
\def\mn@doi{\begingroup\mn@urlcharsother \@ifnextchar [ {\mn@doi@}
  {\mn@doi@[]}}
\def\mn@doi@[#1]#2{\def\@tempa{#1}\ifx\@tempa\@empty \href
  {http://dx.doi.org/#2} {doi:#2}\else \href {http://dx.doi.org/#2} {#1}\fi
  \endgroup}
\def\mn@eprint#1#2{\mn@eprint@#1:#2::\@nil}
\def\mn@eprint@arXiv#1{\href {http://arxiv.org/abs/#1} {{\tt arXiv:#1}}}
\def\mn@eprint@dblp#1{\href {http://dblp.uni-trier.de/rec/bibtex/#1.xml}
  {dblp:#1}}
\def\mn@eprint@#1:#2:#3:#4\@nil{\def\@tempa {#1}\def\@tempb {#2}\def\@tempc
  {#3}\ifx \@tempc \@empty \let \@tempc \@tempb \let \@tempb \@tempa \fi \ifx
  \@tempb \@empty \def\@tempb {arXiv}\fi \@ifundefined
  {mn@eprint@\@tempb}{\@tempb:\@tempc}{\expandafter \expandafter \csname
  mn@eprint@\@tempb\endcsname \expandafter{\@tempc}}}

\bibitem[\protect\citeauthoryear{{Aihara} et~al.,}{{Aihara}
  et~al.}{2018}]{aihara18}
{Aihara} H.,  et~al., 2018, \mn@doi [\pasj] {10.1093/pasj/psx066}, \href
  {https://ui.adsabs.harvard.edu/abs/2018PASJ...70S...4A} {70, S4}

\bibitem[\protect\citeauthoryear{{Aihara} et~al.,}{{Aihara}
  et~al.}{2022}]{aihara22}
{Aihara} H.,  et~al., 2022, \mn@doi [\pasj] {10.1093/pasj/psab122}, \href
  {https://ui.adsabs.harvard.edu/abs/2022PASJ...74..247A} {74, 247}

\bibitem[\protect\citeauthoryear{{Alpaslan} et~al.,}{{Alpaslan}
  et~al.}{2014}]{alpaslan14}
{Alpaslan} M.,  et~al., 2014, \mn@doi [\mnras] {10.1093/mnras/stt2136}, \href
  {https://ui.adsabs.harvard.edu/abs/2014MNRAS.438..177A} {438, 177}

\bibitem[\protect\citeauthoryear{{Bielby} et~al.,}{{Bielby}
  et~al.}{2013}]{bielby13}
{Bielby} R.,  et~al., 2013, \mn@doi [\mnras] {10.1093/mnras/sts639}, \href
  {https://ui.adsabs.harvard.edu/abs/2013MNRAS.430..425B} {430, 425}

\bibitem[\protect\citeauthoryear{{Boquien}, {Burgarella}, {Roehlly}, {Buat},
  {Ciesla}, {Corre}, {Inoue}  \& {Salas}}{{Boquien} et~al.}{2019}]{boquien19}
{Boquien} M.,  {Burgarella} D.,  {Roehlly} Y.,  {Buat} V.,  {Ciesla} L.,
  {Corre} D.,  {Inoue} A.~K.,   {Salas} H.,  2019, \mn@doi [\aap]
  {10.1051/0004-6361/201834156}, \href
  {https://ui.adsabs.harvard.edu/abs/2019A&A...622A.103B} {622, A103}

\bibitem[\protect\citeauthoryear{{Bouwens} et~al.,}{{Bouwens}
  et~al.}{2012}]{bouwens12}
{Bouwens} R.~J.,  et~al., 2012, \mn@doi [\apj] {10.1088/0004-637X/754/2/83},
  \href {https://ui.adsabs.harvard.edu/abs/2012ApJ...754...83B} {754, 83}

\bibitem[\protect\citeauthoryear{{Bouwens}, {Illingworth}, {Ellis}, {Oesch},
  {Paulino-Afonso}, {Ribeiro}  \& {Stefanon}}{{Bouwens}
  et~al.}{2022}]{bouwens22}
{Bouwens} R.~J.,  {Illingworth} G.,  {Ellis} R.~S.,  {Oesch} P.,
  {Paulino-Afonso} A.,  {Ribeiro} B.,   {Stefanon} M.,  2022, \mn@doi [\apj]
  {10.3847/1538-4357/ac618c}, \href
  {https://ui.adsabs.harvard.edu/abs/2022ApJ...931...81B} {931, 81}

\bibitem[\protect\citeauthoryear{{Bruzual} \& {Charlot}}{{Bruzual} \&
  {Charlot}}{2003}]{bruzual03}
{Bruzual} G.,  {Charlot} S.,  2003, \mn@doi [\mnras]
  {10.1046/j.1365-8711.2003.06897.x}, \href
  {https://ui.adsabs.harvard.edu/abs/2003MNRAS.344.1000B} {344, 1000}

\bibitem[\protect\citeauthoryear{{Burgarella}, {Buat}  \&
  {Iglesias-P{\'a}ramo}}{{Burgarella} et~al.}{2005}]{burgarella05}
{Burgarella} D.,  {Buat} V.,   {Iglesias-P{\'a}ramo} J.,  2005, \mn@doi
  [\mnras] {10.1111/j.1365-2966.2005.09131.x}, \href
  {https://ui.adsabs.harvard.edu/abs/2005MNRAS.360.1413B} {360, 1413}

\bibitem[\protect\citeauthoryear{{Calzetti}, {Armus}, {Bohlin}, {Kinney},
  {Koornneef}  \& {Storchi-Bergmann}}{{Calzetti} et~al.}{2000}]{calzetti00}
{Calzetti} D.,  {Armus} L.,  {Bohlin} R.~C.,  {Kinney} A.~L.,  {Koornneef} J.,
   {Storchi-Bergmann} T.,  2000, \mn@doi [\apj] {10.1086/308692}, \href
  {https://ui.adsabs.harvard.edu/abs/2000ApJ...533..682C} {533, 682}

\bibitem[\protect\citeauthoryear{{Capak} et~al.,}{{Capak}
  et~al.}{2011}]{capak11}
{Capak} P.~L.,  et~al., 2011, \mn@doi [\nat] {10.1038/nature09681}, \href
  {https://ui.adsabs.harvard.edu/abs/2011Natur.470..233C} {470, 233}

\bibitem[\protect\citeauthoryear{{Chartab} et~al.,}{{Chartab}
  et~al.}{2021}]{chartab21}
{Chartab} N.,  et~al., 2021, \mn@doi [\apj] {10.3847/1538-4357/abd71f}, \href
  {https://ui.adsabs.harvard.edu/abs/2021ApJ...908..120C} {908, 120}

\bibitem[\protect\citeauthoryear{{Chiang}, {Overzier}  \& {Gebhardt}}{{Chiang}
  et~al.}{2013}]{chiang13}
{Chiang} Y.-K.,  {Overzier} R.,   {Gebhardt} K.,  2013, \mn@doi [\apj]
  {10.1088/0004-637X/779/2/127}, \href
  {https://ui.adsabs.harvard.edu/abs/2013ApJ...779..127C} {779, 127}

\bibitem[\protect\citeauthoryear{{Chiang}, {Overzier}  \& {Gebhardt}}{{Chiang}
  et~al.}{2014}]{chiang14}
{Chiang} Y.-K.,  {Overzier} R.,   {Gebhardt} K.,  2014, \mn@doi [\apjl]
  {10.1088/2041-8205/782/1/L3}, \href
  {https://ui.adsabs.harvard.edu/abs/2014ApJ...782L...3C} {782, L3}

\bibitem[\protect\citeauthoryear{{Chiang}, {Overzier}, {Gebhardt}  \&
  {Henriques}}{{Chiang} et~al.}{2017}]{chiang17}
{Chiang} Y.-K.,  {Overzier} R.~A.,  {Gebhardt} K.,   {Henriques} B.,  2017,
  \mn@doi [\apjl] {10.3847/2041-8213/aa7e7b}, \href
  {https://ui.adsabs.harvard.edu/abs/2017ApJ...844L..23C} {844, L23}

\bibitem[\protect\citeauthoryear{{Clay}, {Thomas}, {Wilkins}  \&
  {Henriques}}{{Clay} et~al.}{2015}]{clay15}
{Clay} S.~J.,  {Thomas} P.~A.,  {Wilkins} S.~M.,   {Henriques} B. M.~B.,  2015,
  \mn@doi [\mnras] {10.1093/mnras/stv818}, \href
  {https://ui.adsabs.harvard.edu/abs/2015MNRAS.451.2692C} {451, 2692}

\bibitem[\protect\citeauthoryear{{Cucciati} et~al.,}{{Cucciati}
  et~al.}{2014}]{cucciati14}
{Cucciati} O.,  et~al., 2014, \mn@doi [\aap] {10.1051/0004-6361/201423811},
  \href {https://ui.adsabs.harvard.edu/abs/2014A&A...570A..16C} {570, A16}

\bibitem[\protect\citeauthoryear{{Cucciati} et~al.,}{{Cucciati}
  et~al.}{2018}]{cucciati18}
{Cucciati} O.,  et~al., 2018, \mn@doi [\aap] {10.1051/0004-6361/201833655},
  \href {https://ui.adsabs.harvard.edu/abs/2018A&A...619A..49C} {619, A49}

\bibitem[\protect\citeauthoryear{{Daddi} et~al.,}{{Daddi}
  et~al.}{2021}]{daddi21}
{Daddi} E.,  et~al., 2021, \mn@doi [\aap] {10.1051/0004-6361/202038700}, \href
  {https://ui.adsabs.harvard.edu/abs/2021A&A...649A..78D} {649, A78}

\bibitem[\protect\citeauthoryear{{Daddi} et~al.,}{{Daddi}
  et~al.}{2022}]{daddi22}
{Daddi} E.,  et~al., 2022, \mn@doi [\apjl] {10.3847/2041-8213/ac531f}, \href
  {https://ui.adsabs.harvard.edu/abs/2022ApJ...926L..21D} {926, L21}

\bibitem[\protect\citeauthoryear{{Desprez} et~al.,}{{Desprez}
  et~al.}{2023}]{desprez23}
{Desprez} G.,  et~al., 2023, \mn@doi [\aap] {10.1051/0004-6361/202243363},
  \href {https://ui.adsabs.harvard.edu/abs/2023A&A...670A..82D} {670, A82}

\bibitem[\protect\citeauthoryear{{Dressler}}{{Dressler}}{1980}]{dressler80}
{Dressler} A.,  1980, \mn@doi [\apj] {10.1086/157753}, \href
  {https://ui.adsabs.harvard.edu/abs/1980ApJ...236..351D} {236, 351}

\bibitem[\protect\citeauthoryear{{Girardi}, {Groenewegen}, {Hatziminaoglou}  \&
  {da Costa}}{{Girardi} et~al.}{2005}]{girardi05}
{Girardi} L.,  {Groenewegen} M.~A.~T.,  {Hatziminaoglou} E.,   {da Costa} L.,
  2005, \mn@doi [\aap] {10.1051/0004-6361:20042352}, \href
  {https://ui.adsabs.harvard.edu/abs/2005A&A...436..895G} {436, 895}

\bibitem[\protect\citeauthoryear{{Harikane} et~al.,}{{Harikane}
  et~al.}{2019}]{harikane19}
{Harikane} Y.,  et~al., 2019, \mn@doi [\apj] {10.3847/1538-4357/ab2cd5}, \href
  {https://ui.adsabs.harvard.edu/abs/2019ApJ...883..142H} {883, 142}

\bibitem[\protect\citeauthoryear{{Harikane} et~al.,}{{Harikane}
  et~al.}{2022}]{harikane22}
{Harikane} Y.,  et~al., 2022, \mn@doi [\apjs] {10.3847/1538-4365/ac3dfc}, \href
  {https://ui.adsabs.harvard.edu/abs/2022ApJS..259...20H} {259, 20}

\bibitem[\protect\citeauthoryear{{Hatch}, {Kurk}, {Pentericci}, {Venemans},
  {Kuiper}, {Miley}  \& {R{\"o}ttgering}}{{Hatch} et~al.}{2011}]{hatch11}
{Hatch} N.~A.,  {Kurk} J.~D.,  {Pentericci} L.,  {Venemans} B.~P.,  {Kuiper}
  E.,  {Miley} G.~K.,   {R{\"o}ttgering} H.~J.~A.,  2011, \mn@doi [\mnras]
  {10.1111/j.1365-2966.2011.18735.x}, \href
  {https://ui.adsabs.harvard.edu/abs/2011MNRAS.415.2993H} {415, 2993}

\bibitem[\protect\citeauthoryear{{Hatch} et~al.,}{{Hatch}
  et~al.}{2014}]{hatch14}
{Hatch} N.~A.,  et~al., 2014, \mn@doi [\mnras] {10.1093/mnras/stu1725}, \href
  {https://ui.adsabs.harvard.edu/abs/2014MNRAS.445..280H} {445, 280}

\bibitem[\protect\citeauthoryear{{Henriques}, {White}, {Lemson}, {Thomas},
  {Guo}, {Marleau}  \& {Overzier}}{{Henriques} et~al.}{2012}]{henriques12}
{Henriques} B. M.~B.,  {White} S. D.~M.,  {Lemson} G.,  {Thomas} P.~A.,  {Guo}
  Q.,  {Marleau} G.-D.,   {Overzier} R.~A.,  2012, \mn@doi [\mnras]
  {10.1111/j.1365-2966.2012.20521.x}, \href
  {https://ui.adsabs.harvard.edu/abs/2012MNRAS.421.2904H} {421, 2904}

\bibitem[\protect\citeauthoryear{{Henriques}, {White}, {Thomas}, {Angulo},
  {Guo}, {Lemson}, {Springel}  \& {Overzier}}{{Henriques}
  et~al.}{2015}]{henriques15}
{Henriques} B. M.~B.,  {White} S. D.~M.,  {Thomas} P.~A.,  {Angulo} R.,  {Guo}
  Q.,  {Lemson} G.,  {Springel} V.,   {Overzier} R.,  2015, \mn@doi [\mnras]
  {10.1093/mnras/stv705}, \href
  {https://ui.adsabs.harvard.edu/abs/2015MNRAS.451.2663H} {451, 2663}

\bibitem[\protect\citeauthoryear{{Higuchi} et~al.,}{{Higuchi}
  et~al.}{2019}]{higuchi19}
{Higuchi} R.,  et~al., 2019, \mn@doi [\apj] {10.3847/1538-4357/ab2192}, \href
  {https://ui.adsabs.harvard.edu/abs/2019ApJ...879...28H} {879, 28}

\bibitem[\protect\citeauthoryear{{Hu} et~al.,}{{Hu} et~al.}{2021}]{hu21}
{Hu} W.,  et~al., 2021, \mn@doi [Nature Astronomy]
  {10.1038/s41550-020-01291-y}, \href
  {https://ui.adsabs.harvard.edu/abs/2021NatAs...5..485H} {5, 485}

\bibitem[\protect\citeauthoryear{{Ito} et~al.,}{{Ito} et~al.}{2019}]{ito19}
{Ito} K.,  et~al., 2019, \mn@doi [\apj] {10.3847/1538-4357/ab1f0c}, \href
  {https://ui.adsabs.harvard.edu/abs/2019ApJ...878...68I} {878, 68}

\bibitem[\protect\citeauthoryear{{Ito} et~al.,}{{Ito} et~al.}{2020}]{ito20}
{Ito} K.,  et~al., 2020, \mn@doi [\apj] {10.3847/1538-4357/aba269}, \href
  {https://ui.adsabs.harvard.edu/abs/2020ApJ...899....5I} {899, 5}

\bibitem[\protect\citeauthoryear{{Kennicutt}}{{Kennicutt}}{1998}]{kennicutt98}
{Kennicutt} Robert~C. J.,  1998, \mn@doi [\araa]
  {10.1146/annurev.astro.36.1.189}, \href
  {https://ui.adsabs.harvard.edu/abs/1998ARA&A..36..189K} {36, 189}

\bibitem[\protect\citeauthoryear{{Koyama} et~al.,}{{Koyama}
  et~al.}{2013}]{koyama13}
{Koyama} Y.,  et~al., 2013, \mn@doi [\mnras] {10.1093/mnras/stt1035}, \href
  {https://ui.adsabs.harvard.edu/abs/2013MNRAS.434..423K} {434, 423}

\bibitem[\protect\citeauthoryear{{Kubo} et~al.,}{{Kubo} et~al.}{2019}]{kubo19}
{Kubo} M.,  et~al., 2019, \mn@doi [\apj] {10.3847/1538-4357/ab5a80}, \href
  {https://ui.adsabs.harvard.edu/abs/2019ApJ...887..214K} {887, 214}

\bibitem[\protect\citeauthoryear{{Kubo} et~al.,}{{Kubo} et~al.}{2021}]{kubo21}
{Kubo} M.,  et~al., 2021, \mn@doi [\apj] {10.3847/1538-4357/ac0cf8}, \href
  {https://ui.adsabs.harvard.edu/abs/2021ApJ...919....6K} {919, 6}

\bibitem[\protect\citeauthoryear{{Laigle} et~al.,}{{Laigle}
  et~al.}{2016}]{laigle16}
{Laigle} C.,  et~al., 2016, \mn@doi [\apjs] {10.3847/0067-0049/224/2/24}, \href
  {https://ui.adsabs.harvard.edu/abs/2016ApJS..224...24L} {224, 24}

\bibitem[\protect\citeauthoryear{{Laporte}, {Zitrin}, {Dole},
  {Roberts-Borsani}, {Furtak}  \& {Witten}}{{Laporte} et~al.}{2022}]{laporte22}
{Laporte} N.,  {Zitrin} A.,  {Dole} H.,  {Roberts-Borsani} G.,  {Furtak} L.~J.,
    {Witten} C.,  2022, arXiv e-prints, \href
  {https://ui.adsabs.harvard.edu/abs/2022arXiv220804930L} {p. arXiv:2208.04930}

\bibitem[\protect\citeauthoryear{{Lawrence} et~al.,}{{Lawrence}
  et~al.}{2007}]{lawrence07}
{Lawrence} A.,  et~al., 2007, \mn@doi [\mnras]
  {10.1111/j.1365-2966.2007.12040.x}, \href
  {https://ui.adsabs.harvard.edu/abs/2007MNRAS.379.1599L} {379, 1599}

\bibitem[\protect\citeauthoryear{{Le F{\`e}vre} et~al.,}{{Le F{\`e}vre}
  et~al.}{2015}]{lefevre15}
{Le F{\`e}vre} O.,  et~al., 2015, \mn@doi [\aap] {10.1051/0004-6361/201423829},
  \href {https://ui.adsabs.harvard.edu/abs/2015A&A...576A..79L} {576, A79}

\bibitem[\protect\citeauthoryear{{Lehmer} et~al.,}{{Lehmer}
  et~al.}{2013}]{lehmer13}
{Lehmer} B.~D.,  et~al., 2013, \mn@doi [\apj] {10.1088/0004-637X/765/2/87},
  \href {https://ui.adsabs.harvard.edu/abs/2013ApJ...765...87L} {765, 87}

\bibitem[\protect\citeauthoryear{{Lemaux} et~al.,}{{Lemaux}
  et~al.}{2014}]{lemaux14}
{Lemaux} B.~C.,  et~al., 2014, \mn@doi [\aap] {10.1051/0004-6361/201423828},
  \href {https://ui.adsabs.harvard.edu/abs/2014A&A...572A..41L} {572, A41}

\bibitem[\protect\citeauthoryear{{Lemaux} et~al.,}{{Lemaux}
  et~al.}{2018}]{lemaux18}
{Lemaux} B.~C.,  et~al., 2018, \mn@doi [\aap] {10.1051/0004-6361/201730870},
  \href {https://ui.adsabs.harvard.edu/abs/2018A&A...615A..77L} {615, A77}

\bibitem[\protect\citeauthoryear{{Lemaux} et~al.,}{{Lemaux}
  et~al.}{2022}]{lemaux22}
{Lemaux} B.~C.,  et~al., 2022, \mn@doi [\aap] {10.1051/0004-6361/202039346},
  \href {https://ui.adsabs.harvard.edu/abs/2022A&A...662A..33L} {662, A33}

\bibitem[\protect\citeauthoryear{{Lim}, {Scott}, {Babul}, {Barnes}, {Kay},
  {McCarthy}, {Rennehan}  \& {Vogelsberger}}{{Lim} et~al.}{2021}]{lim21}
{Lim} S.,  {Scott} D.,  {Babul} A.,  {Barnes} D.~J.,  {Kay} S.~T.,  {McCarthy}
  I.~G.,  {Rennehan} D.,   {Vogelsberger} M.,  2021, \mn@doi [\mnras]
  {10.1093/mnras/staa3693}, \href
  {https://ui.adsabs.harvard.edu/abs/2021MNRAS.501.1803L} {501, 1803}

\bibitem[\protect\citeauthoryear{{Lovell}, {Thomas}  \& {Wilkins}}{{Lovell}
  et~al.}{2018}]{lovell18}
{Lovell} C.~C.,  {Thomas} P.~A.,   {Wilkins} S.~M.,  2018, \mn@doi [\mnras]
  {10.1093/mnras/stx3090}, \href
  {https://ui.adsabs.harvard.edu/abs/2018MNRAS.474.4612L} {474, 4612}

\bibitem[\protect\citeauthoryear{{Macuga} et~al.,}{{Macuga}
  et~al.}{2019}]{macuga19}
{Macuga} M.,  et~al., 2019, \mn@doi [\apj] {10.3847/1538-4357/ab0746}, \href
  {https://ui.adsabs.harvard.edu/abs/2019ApJ...874...54M} {874, 54}

\bibitem[\protect\citeauthoryear{{Malavasi}, {Lee}, {Dey}, {Xue}, {Huang}  \&
  {Shi}}{{Malavasi} et~al.}{2021}]{malavasi21}
{Malavasi} N.,  {Lee} K.-S.,  {Dey} A.,  {Xue} R.,  {Huang} Y.,   {Shi} K.,
  2021, \mn@doi [\apj] {10.3847/1538-4357/ac1c6e}, \href
  {https://ui.adsabs.harvard.edu/abs/2021ApJ...921..103M} {921, 103}

\bibitem[\protect\citeauthoryear{{Marinello} et~al.,}{{Marinello}
  et~al.}{2020}]{marinello20}
{Marinello} M.,  et~al., 2020, \mn@doi [\mnras] {10.1093/mnras/stz3333}, \href
  {https://ui.adsabs.harvard.edu/abs/2020MNRAS.492.1991M} {492, 1991}

\bibitem[\protect\citeauthoryear{{McConachie} et~al.,}{{McConachie}
  et~al.}{2022}]{mcconachie22}
{McConachie} I.,  et~al., 2022, \mn@doi [\apj] {10.3847/1538-4357/ac2b9f},
  \href {https://ui.adsabs.harvard.edu/abs/2022ApJ...926...37M} {926, 37}

\bibitem[\protect\citeauthoryear{{McCracken} et~al.,}{{McCracken}
  et~al.}{2012}]{mccracken12}
{McCracken} H.~J.,  et~al., 2012, \mn@doi [\aap] {10.1051/0004-6361/201219507},
  \href {https://ui.adsabs.harvard.edu/abs/2012A&A...544A.156M} {544, A156}

\bibitem[\protect\citeauthoryear{{Meiksin}}{{Meiksin}}{2006}]{meiksin06}
{Meiksin} A.,  2006, \mn@doi [\mnras] {10.1111/j.1365-2966.2005.09756.x}, \href
  {https://ui.adsabs.harvard.edu/abs/2006MNRAS.365..807M} {365, 807}

\bibitem[\protect\citeauthoryear{{Miller} et~al.,}{{Miller}
  et~al.}{2018}]{miller18}
{Miller} T.~B.,  et~al., 2018, \mn@doi [\nat] {10.1038/s41586-018-0025-2},
  \href {https://ui.adsabs.harvard.edu/abs/2018Natur.556..469M} {556, 469}

\bibitem[\protect\citeauthoryear{{Morishita} et~al.,}{{Morishita}
  et~al.}{2022}]{morishita22}
{Morishita} T.,  et~al., 2022, arXiv e-prints, \href
  {https://ui.adsabs.harvard.edu/abs/2022arXiv221109097M} {p. arXiv:2211.09097}

\bibitem[\protect\citeauthoryear{{Muldrew}, {Hatch}  \& {Cooke}}{{Muldrew}
  et~al.}{2015}]{muldrew15}
{Muldrew} S.~I.,  {Hatch} N.~A.,   {Cooke} E.~A.,  2015, \mn@doi [\mnras]
  {10.1093/mnras/stv1449}, \href
  {https://ui.adsabs.harvard.edu/abs/2015MNRAS.452.2528M} {452, 2528}

\bibitem[\protect\citeauthoryear{{Noirot} et~al.,}{{Noirot}
  et~al.}{2018}]{noirot18}
{Noirot} G.,  et~al., 2018, \mn@doi [\apj] {10.3847/1538-4357/aabadb}, \href
  {https://ui.adsabs.harvard.edu/abs/2018ApJ...859...38N} {859, 38}

\bibitem[\protect\citeauthoryear{{Noll}, {Burgarella}, {Giovannoli}, {Buat},
  {Marcillac}  \& {Mu{\~n}oz-Mateos}}{{Noll} et~al.}{2009}]{noll09}
{Noll} S.,  {Burgarella} D.,  {Giovannoli} E.,  {Buat} V.,  {Marcillac} D.,
  {Mu{\~n}oz-Mateos} J.~C.,  2009, \mn@doi [\aap]
  {10.1051/0004-6361/200912497}, \href
  {https://ui.adsabs.harvard.edu/abs/2009A&A...507.1793N} {507, 1793}

\bibitem[\protect\citeauthoryear{{Ono} et~al.,}{{Ono} et~al.}{2018}]{ono18}
{Ono} Y.,  et~al., 2018, \mn@doi [\pasj] {10.1093/pasj/psx103}, \href
  {https://ui.adsabs.harvard.edu/abs/2018PASJ...70S..10O} {70, S10}

\bibitem[\protect\citeauthoryear{{Onoue} et~al.,}{{Onoue}
  et~al.}{2018}]{onoue18}
{Onoue} M.,  et~al., 2018, \mn@doi [\pasj] {10.1093/pasj/psx092}, \href
  {https://ui.adsabs.harvard.edu/abs/2018PASJ...70S..31O} {70, S31}

\bibitem[\protect\citeauthoryear{{Oteo} et~al.,}{{Oteo} et~al.}{2018}]{oteo18}
{Oteo} I.,  et~al., 2018, \mn@doi [\apj] {10.3847/1538-4357/aaa1f1}, \href
  {https://ui.adsabs.harvard.edu/abs/2018ApJ...856...72O} {856, 72}

\bibitem[\protect\citeauthoryear{{Ouchi} et~al.,}{{Ouchi}
  et~al.}{2005}]{ouchi05}
{Ouchi} M.,  et~al., 2005, \mn@doi [\apjl] {10.1086/428499}, \href
  {https://ui.adsabs.harvard.edu/abs/2005ApJ...620L...1O} {620, L1}

\bibitem[\protect\citeauthoryear{{Pascarelle}, {Windhorst}  \&
  {Keel}}{{Pascarelle} et~al.}{1998}]{pascarelle98}
{Pascarelle} S.~M.,  {Windhorst} R.~A.,   {Keel} W.~C.,  1998, \mn@doi [\aj]
  {10.1086/300634}, \href
  {https://ui.adsabs.harvard.edu/abs/1998AJ....116.2659P} {116, 2659}

\bibitem[\protect\citeauthoryear{{Peebles}}{{Peebles}}{1980}]{peebles80}
{Peebles} P.~J.~E.,  1980, {The large-scale structure of the universe}

\bibitem[\protect\citeauthoryear{{Pentericci} et~al.,}{{Pentericci}
  et~al.}{2000}]{pentericci00}
{Pentericci} L.,  et~al., 2000, \aap, \href
  {https://ui.adsabs.harvard.edu/abs/2000A&A...361L..25P} {361, L25}

\bibitem[\protect\citeauthoryear{{Pozzetti} et~al.,}{{Pozzetti}
  et~al.}{2007}]{pozzetti07}
{Pozzetti} L.,  et~al., 2007, \mn@doi [\aap] {10.1051/0004-6361:20077609},
  \href {https://ui.adsabs.harvard.edu/abs/2007A&A...474..443P} {474, 443}

\bibitem[\protect\citeauthoryear{{Remus}, {Dolag}  \& {Dannerbauer}}{{Remus}
  et~al.}{2022}]{remus22}
{Remus} R.-S.,  {Dolag} K.,   {Dannerbauer} H.,  2022, arXiv e-prints, \href
  {https://ui.adsabs.harvard.edu/abs/2022arXiv220801053R} {p. arXiv:2208.01053}

\bibitem[\protect\citeauthoryear{{Rotermund} et~al.,}{{Rotermund}
  et~al.}{2021}]{rotermund21}
{Rotermund} K.~M.,  et~al., 2021, \mn@doi [\mnras] {10.1093/mnras/stab103},
  \href {https://ui.adsabs.harvard.edu/abs/2021MNRAS.502.1797R} {502, 1797}

\bibitem[\protect\citeauthoryear{{Salpeter}}{{Salpeter}}{1955}]{salpeter55}
{Salpeter} E.~E.,  1955, \mn@doi [\apj] {10.1086/145971}, \href
  {https://ui.adsabs.harvard.edu/abs/1955ApJ...121..161S} {121, 161}

\bibitem[\protect\citeauthoryear{{Sattari}, {Mobasher}, {Chartab}, {Darvish},
  {Shivaei}, {Scoville}  \& {Sobral}}{{Sattari} et~al.}{2021}]{sattari21}
{Sattari} Z.,  {Mobasher} B.,  {Chartab} N.,  {Darvish} B.,  {Shivaei} I.,
  {Scoville} N.,   {Sobral} D.,  2021, \mn@doi [\apj]
  {10.3847/1538-4357/abe5a3}, \href
  {https://ui.adsabs.harvard.edu/abs/2021ApJ...910...57S} {910, 57}

\bibitem[\protect\citeauthoryear{{Sawicki} et~al.,}{{Sawicki}
  et~al.}{2019}]{sawicki19}
{Sawicki} M.,  et~al., 2019, \mn@doi [\mnras] {10.1093/mnras/stz2522}, \href
  {https://ui.adsabs.harvard.edu/abs/2019MNRAS.489.5202S} {489, 5202}

\bibitem[\protect\citeauthoryear{{Sawicki}, {Arcila-Osejo}, {Golob}, {Moutard},
  {Arnouts}  \& {Cheema}}{{Sawicki} et~al.}{2020}]{sawicki20}
{Sawicki} M.,  {Arcila-Osejo} L.,  {Golob} A.,  {Moutard} T.,  {Arnouts} S.,
  {Cheema} G.~K.,  2020, \mn@doi [\mnras] {10.1093/mnras/staa779}, \href
  {https://ui.adsabs.harvard.edu/abs/2020MNRAS.494.1366S} {494, 1366}

\bibitem[\protect\citeauthoryear{{Shamshiri}, {Thomas}, {Henriques}, {Tojeiro},
  {Lemson}, {Oliver}  \& {Wilkins}}{{Shamshiri} et~al.}{2015}]{shamshiri15}
{Shamshiri} S.,  {Thomas} P.~A.,  {Henriques} B.~M.,  {Tojeiro} R.,  {Lemson}
  G.,  {Oliver} S.~J.,   {Wilkins} S.,  2015, \mn@doi [\mnras]
  {10.1093/mnras/stv883}, \href
  {https://ui.adsabs.harvard.edu/abs/2015MNRAS.451.2681S} {451, 2681}

\bibitem[\protect\citeauthoryear{{Shi} et~al.,}{{Shi} et~al.}{2019}]{shi19}
{Shi} K.,  et~al., 2019, \mn@doi [\apj] {10.3847/1538-4357/ab2118}, \href
  {https://ui.adsabs.harvard.edu/abs/2019ApJ...879....9S} {879, 9}

\bibitem[\protect\citeauthoryear{{Shi}, {Toshikawa}, {Cai}, {Lee}  \&
  {Fang}}{{Shi} et~al.}{2020}]{shi20}
{Shi} K.,  {Toshikawa} J.,  {Cai} Z.,  {Lee} K.-S.,   {Fang} T.,  2020, \mn@doi
  [\apj] {10.3847/1538-4357/aba626}, \href
  {https://ui.adsabs.harvard.edu/abs/2020ApJ...899...79S} {899, 79}

\bibitem[\protect\citeauthoryear{{Shi}, {Toshikawa}, {Lee}, {Wang}, {Cai}  \&
  {Fang}}{{Shi} et~al.}{2021}]{shi21}
{Shi} K.,  {Toshikawa} J.,  {Lee} K.-S.,  {Wang} T.,  {Cai} Z.,   {Fang} T.,
  2021, \mn@doi [\apj] {10.3847/1538-4357/abe62e}, \href
  {https://ui.adsabs.harvard.edu/abs/2021ApJ...911...46S} {911, 46}

\bibitem[\protect\citeauthoryear{{Shimakawa}, {Kodama}, {Tadaki}, {Hayashi},
  {Koyama}  \& {Tanaka}}{{Shimakawa} et~al.}{2015}]{shimakawa15}
{Shimakawa} R.,  {Kodama} T.,  {Tadaki} K.-i.,  {Hayashi} M.,  {Koyama} Y.,
  {Tanaka} I.,  2015, \mn@doi [\mnras] {10.1093/mnras/stv051}, \href
  {https://ui.adsabs.harvard.edu/abs/2015MNRAS.448..666S} {448, 666}

\bibitem[\protect\citeauthoryear{{Shimasaku} et~al.,}{{Shimasaku}
  et~al.}{2003}]{shimasaku03}
{Shimasaku} K.,  et~al., 2003, \mn@doi [\apjl] {10.1086/374880}, \href
  {https://ui.adsabs.harvard.edu/abs/2003ApJ...586L.111S} {586, L111}

\bibitem[\protect\citeauthoryear{{Sillassen} et~al.,}{{Sillassen}
  et~al.}{2022}]{sillassen22}
{Sillassen} N.~B.,  et~al., 2022, \mn@doi [\aap] {10.1051/0004-6361/202244661},
  \href {https://ui.adsabs.harvard.edu/abs/2022A&A...665L...7S} {665, L7}

\bibitem[\protect\citeauthoryear{{Steidel} \& {Hamilton}}{{Steidel} \&
  {Hamilton}}{1992}]{steidel92}
{Steidel} C.~C.,  {Hamilton} D.,  1992, \mn@doi [\aj] {10.1086/116287}, \href
  {https://ui.adsabs.harvard.edu/abs/1992AJ....104..941S} {104, 941}

\bibitem[\protect\citeauthoryear{{Steidel}, {Adelberger}, {Dickinson},
  {Giavalisco}, {Pettini}  \& {Kellogg}}{{Steidel} et~al.}{1998}]{steidel98}
{Steidel} C.~C.,  {Adelberger} K.~L.,  {Dickinson} M.,  {Giavalisco} M.,
  {Pettini} M.,   {Kellogg} M.,  1998, \mn@doi [\apj] {10.1086/305073}, \href
  {https://ui.adsabs.harvard.edu/abs/1998ApJ...492..428S} {492, 428}

\bibitem[\protect\citeauthoryear{{Tanaka}}{{Tanaka}}{2015}]{tanaka15}
{Tanaka} M.,  2015, \mn@doi [\apj] {10.1088/0004-637X/801/1/20}, \href
  {https://ui.adsabs.harvard.edu/abs/2015ApJ...801...20T} {801, 20}

\bibitem[\protect\citeauthoryear{{Tanaka} et~al.,}{{Tanaka}
  et~al.}{2018}]{tanaka18}
{Tanaka} M.,  et~al., 2018, \mn@doi [\pasj] {10.1093/pasj/psx077}, \href
  {https://ui.adsabs.harvard.edu/abs/2018PASJ...70S...9T} {70, S9}

\bibitem[\protect\citeauthoryear{{Thomas}, {Maraston}, {Bender}  \& {Mendes de
  Oliveira}}{{Thomas} et~al.}{2005}]{thomas05}
{Thomas} D.,  {Maraston} C.,  {Bender} R.,   {Mendes de Oliveira} C.,  2005,
  \mn@doi [\apj] {10.1086/426932}, \href
  {https://ui.adsabs.harvard.edu/abs/2005ApJ...621..673T} {621, 673}

\bibitem[\protect\citeauthoryear{{Toshikawa} et~al.,}{{Toshikawa}
  et~al.}{2012}]{toshikawa12}
{Toshikawa} J.,  et~al., 2012, \mn@doi [\apj] {10.1088/0004-637X/750/2/137},
  \href {https://ui.adsabs.harvard.edu/abs/2012ApJ...750..137T} {750, 137}

\bibitem[\protect\citeauthoryear{{Toshikawa} et~al.,}{{Toshikawa}
  et~al.}{2016}]{toshikawa16}
{Toshikawa} J.,  et~al., 2016, \mn@doi [\apj] {10.3847/0004-637X/826/2/114},
  \href {https://ui.adsabs.harvard.edu/abs/2016ApJ...826..114T} {826, 114}

\bibitem[\protect\citeauthoryear{{Toshikawa} et~al.,}{{Toshikawa}
  et~al.}{2018}]{toshikawa18}
{Toshikawa} J.,  et~al., 2018, \mn@doi [\pasj] {10.1093/pasj/psx102}, \href
  {https://ui.adsabs.harvard.edu/abs/2018PASJ...70S..12T} {70, S12}

\bibitem[\protect\citeauthoryear{{Toshikawa}, {Malkan}, {Kashikawa},
  {Overzier}, {Uchiyama}, {Ota}, {Ishikawa}  \& {Ito}}{{Toshikawa}
  et~al.}{2020}]{toshikawa20}
{Toshikawa} J.,  {Malkan} M.~A.,  {Kashikawa} N.,  {Overzier} R.,  {Uchiyama}
  H.,  {Ota} K.,  {Ishikawa} S.,   {Ito} K.,  2020, \mn@doi [\apj]
  {10.3847/1538-4357/ab5e85}, \href
  {https://ui.adsabs.harvard.edu/abs/2020ApJ...888...89T} {888, 89}

\bibitem[\protect\citeauthoryear{{Trudeau}, {Willis}, {Rennehan}, {Canning},
  {Carnall}, {Poggianti}, {Noordeh}  \& {Pierre}}{{Trudeau}
  et~al.}{2022}]{trudeau22}
{Trudeau} A.,  {Willis} J.~P.,  {Rennehan} D.,  {Canning} R.~E.~A.,  {Carnall}
  A.~C.,  {Poggianti} B.,  {Noordeh} E.,   {Pierre} M.,  2022, \mn@doi [\mnras]
  {10.1093/mnras/stac1760}, \href
  {https://ui.adsabs.harvard.edu/abs/2022MNRAS.515.2529T} {515, 2529}

\bibitem[\protect\citeauthoryear{{Uchiyama} et~al.,}{{Uchiyama}
  et~al.}{2018}]{uchiyama18}
{Uchiyama} H.,  et~al., 2018, \mn@doi [\pasj] {10.1093/pasj/psx112}, \href
  {https://ui.adsabs.harvard.edu/abs/2018PASJ...70S..32U} {70, S32}

\bibitem[\protect\citeauthoryear{{Venemans} et~al.,}{{Venemans}
  et~al.}{2007}]{venemans07}
{Venemans} B.~P.,  et~al., 2007, \mn@doi [\aap] {10.1051/0004-6361:20053941},
  \href {https://ui.adsabs.harvard.edu/abs/2007A&A...461..823V} {461, 823}

\bibitem[\protect\citeauthoryear{{Vito} et~al.,}{{Vito} et~al.}{2020}]{vito20}
{Vito} F.,  et~al., 2020, \mn@doi [\aap] {10.1051/0004-6361/202038848}, \href
  {https://ui.adsabs.harvard.edu/abs/2020A&A...642A.149V} {642, A149}

\bibitem[\protect\citeauthoryear{{Wang} et~al.,}{{Wang} et~al.}{2016}]{wang16}
{Wang} T.,  et~al., 2016, \mn@doi [\apj] {10.3847/0004-637X/828/1/56}, \href
  {https://ui.adsabs.harvard.edu/abs/2016ApJ...828...56W} {828, 56}

\bibitem[\protect\citeauthoryear{{Weaver} et~al.,}{{Weaver}
  et~al.}{2022}]{weaver22}
{Weaver} J.~R.,  et~al., 2022, \mn@doi [\apjs] {10.3847/1538-4365/ac3078},
  \href {https://ui.adsabs.harvard.edu/abs/2022ApJS..258...11W} {258, 11}

\bibitem[\protect\citeauthoryear{{Wylezalek} et~al.,}{{Wylezalek}
  et~al.}{2013}]{wylezalek13}
{Wylezalek} D.,  et~al., 2013, \mn@doi [\apj] {10.1088/0004-637X/769/1/79},
  \href {https://ui.adsabs.harvard.edu/abs/2013ApJ...769...79W} {769, 79}

\bibitem[\protect\citeauthoryear{{Zhang} et~al.,}{{Zhang}
  et~al.}{2022}]{zhang22}
{Zhang} J.,  et~al., 2022, \mn@doi [\mnras] {10.1093/mnras/stac1083}, \href
  {https://ui.adsabs.harvard.edu/abs/2022MNRAS.513.4814Z} {513, 4814}

\bibitem[\protect\citeauthoryear{{van der Burg}, {Hildebrandt}  \&
  {Erben}}{{van der Burg} et~al.}{2010}]{burg10}
{van der Burg} R.~F.~J.,  {Hildebrandt} H.,   {Erben} T.,  2010, \mn@doi [\aap]
  {10.1051/0004-6361/200913812}, \href
  {https://ui.adsabs.harvard.edu/abs/2010A&A...523A..74V} {523, A74}

\makeatother
\end{thebibliography}

\appendix
\section{Comparison with phot-$z$ catalogues} \label{sec:comp}

\begin{figure*}
\includegraphics[width=\textwidth,bb=0 0 1080 367]{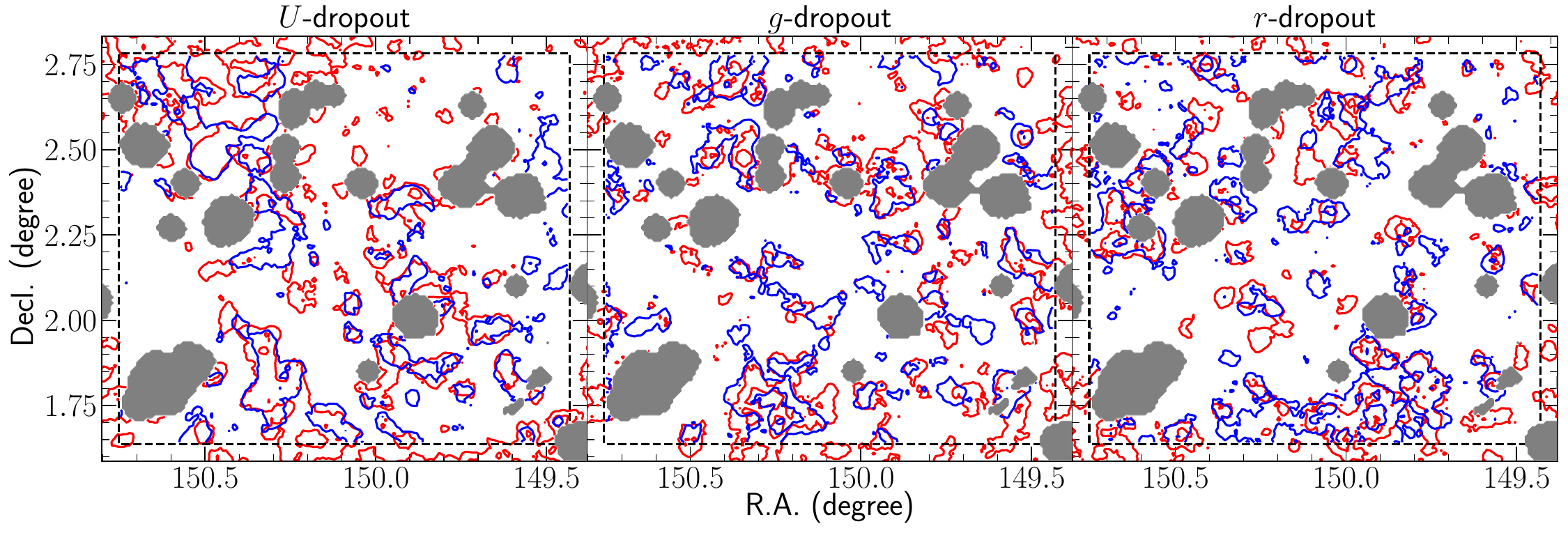}
\caption{Overdensity contours of $U$-, $g$-, and $r$-dropout galaxies based on the HSC-SSP dropout catalogues
    (red) and COSMOS2020 CLASSIC catalogue with EAZY photometric redshifts (blue).
    The gray regions are the masked areas of either the HSC-SSP or COSMOS2020, and the dashed lines indicate the
    FoV of the COSMOS2020.}
\label{fig:comp_cntr}
\end{figure*}

\begin{figure*}
\includegraphics[width=\textwidth,bb=0 0 1080 367]{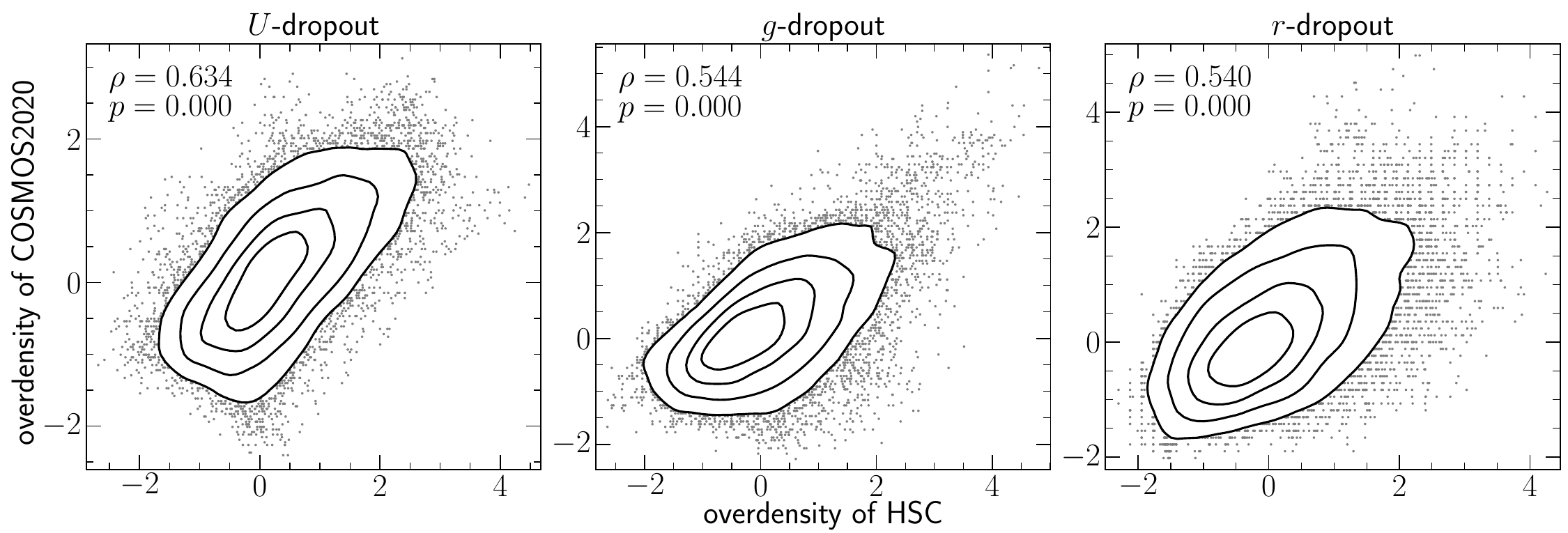}
\caption{Comparison between overdensity measured by HSC's dropout and COSMOS2020's phot-$z$ catalogues on the same
    sky area.
    The black solid lines indicate contours including 25\%, 50\%, 75\%, and 90\% of the data points.
    Spearman's rank correlation test confirmed that overdensity maps created by both catalogues are statistically
    consistent, and its correlation coefficient, $\rho$, and $p$-value are indicated at the upper left corner in
    each panel.}
\label{fig:comp_over}
\end{figure*}

\begin{figure*}
\includegraphics[width=\textwidth,bb=0 0 885 360]{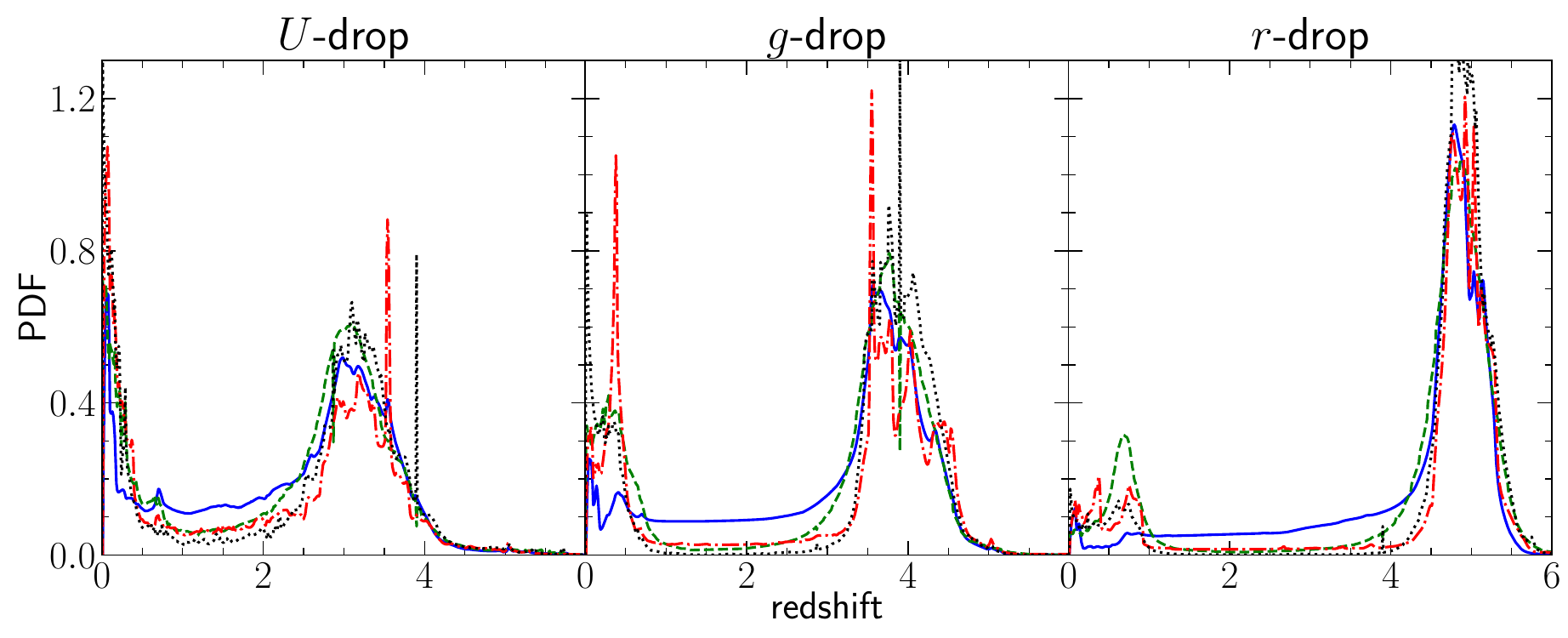}
\caption{Average phot-$z$ PDFs for our dropout samples derived from the four types of
    COSMOS2020 phot-z catalogues, exploiting different algorithms for photometry (CLASSIC/FARMER) and phot-$z$
    estimation (EAZY/LePhare).
    The blue, green, red, and black lines are derived from CLASSIC with EAZY, CLASSIC with LePhare, FARMER with
    EAZY, and FARMER with LePhare, respectively.
    The derived phot-$z$ PDFs vary modestly depending on the type of photometric catalogue or phot-$z$ code used.
    Especially, there are some spurious spikes.}
\label{fig:zPDF_4type}
\end{figure*}

Our protocluster search is based on dropout galaxies, which represent young, star-forming galaxies and contain a
certain fraction of contaminations.
It has the merit of being a well established technique that can be homogeneously applied across a large survey
area.
We check the reliability of our results by directly comparing with overdensity maps based on another galaxy
sample, covering 6\% of the survey area exploited in this study.
Specifically, we use the COSMOS2020 catalogue \citep{weaver22} and select galaxies having
$2.7<z_\mathrm{phot}<3.5$, $3.4<z_\mathrm{phot}<4.2$, and $4.4<z_\mathrm{phot}<5.2$ as the counterparts of $U$-,
$g$-, and $r$-dropout galaxies, respectively.
Then, their overdensity is calculated as we did for our dropout samples.
As shown in Figure~\ref{fig:comp_cntr} and Figure~\ref{fig:comp_over}, the overdensity maps based on both
catalogues are found to be consistent.
Overdensity in both maps are significantly correlated ($p\ll0.01$ by Spearmann's rank correlation test).
These consistent results can be found in all flavours of the COSMOS2020 catalogue (CLASSIC or FARMER catalogues
with EAZY or LePhare phot-$z$ codes) and all three redshift samples.
It should be noted that there is some inconsistency even among four flavours of the COSMOS2020 catalogue
(Figure~\ref{fig:zPDF_4type}).

In addition, the same analysis is performed on the phot-$z$ catalogue constructed by \citet{desprez23}, which
covers a whole DUD layer but is based on the HSC-SSP dataset.
We have confirmed that the overdensity maps of dropout galaxies are consistent ($p\ll0.01$) even with those of the
phot-$z$ galaxies of \citet{desprez23}.
The strength of correlation is found to be $\rho=0.65\mathrm{-}0.75$, which is higher than the case of COSMOS2020
catalogue ($\rho=0.54\mathrm{-}0.64$).
This would be attributed to that the phot-$z$ galaxies of \citet{desprez23} is also based on the same imaging
dataset and source extraction/photometry pipeline as our dropout samples.

Although the phot-$z$ catalogue of \citet{desprez23} may be less independent from our dropout catalogue as both
catalogues utilise the same dataset,
we find no evidence for the overdensity maps to be seriously affected by contaminations or artificial failures in
the HSC-SSP dataset.
The results of our sanity check boost confidence in the reliability of our catalogue of protocluster candidates
at $z\sim3\mathrm{-}5$.

\section{Comparison with known protoclusters from literature} \label{sec:lite}

As the footprint and redshift range of our protocluster search cover some known protoclusters, we list such
protoclusters and their overdensity measured by this study in Table~\ref{tab:lite}.
It is worth emphasizing that the definition of protoclusters or the method of protocluster identification varies
largely across the literature.
We selected protoclusters in which at least three member galaxies are spectroscopically confirmed or in which
there are a large number of member candidates with phot-$z$. 
Some protoclusters were discovered by blind spectroscopic surveys, which can identify small protoclusters without 
foreground/background contaminations.
In addition, the bias of the specific galaxy population traced may have a large impact on the protocluster search
\citep[e.g.,][]{shi19}.
For example, \citet{mcconachie22} confirmed a protocluster at $z=3.3665$ in which 73\% of member galaxies are
quiescent at $M_{\ast}>10^{11}\,\mathrm{M_{\sun}}$.
It is difficult or impossible to identify such protoclusters by our protocluster search, which is based on the
surface overdensity of dropout galaxies.
In spite of the variety of protocluster identification techniques, the overdensity at the positions of known
protoclusters from the literature is significantly biased toward higher densities.
The mean (median) dropout overdensity at their location is 1.9 (1.7)$\sigma$, whereas the typical overdensity
measurement within a randomly placed aperture would by construction be $\sim 0$.
That said, we recognise that despite their above-average overdensity measurement, the known overdense structures
do not formally qualify as protocluster candidates per our working definition, as their peak overdensity of
dropout galaxies within a $0.75\,\mathrm{pMpc}$ aperture does not exceed $4\sigma$.
This iterates the previously raised caveat of completeness, which is perhaps most clearly documented in Figure~7
of \citet{toshikawa16}, namely that many high-redshift galaxies that will end up in massive galaxy clusters by
$z=0$ (and that therefore by definition qualify as protocluster galaxies) do not reside in regions of $>4\sigma$
sky overdensity of dropout galaxies.

\begin{table*}
\caption{Overdensity of dropout galaxies at the location of known protoclusters}
\label{tab:lite}
\begin{tabular}{cccccc}
\hline
Ref. & R.A. (deg) & Decl. (deg) & redshift & overdensity & type$^\text{a}$ \\
\hline
\citet{capak11} & 150.086 & 2.589 & 5.298 & 1.9 & targeted -- SMG \\
\citet{toshikawa20} & 36.190 & -4.930 & 4.898 & 3.0 & blind -- LBG \\
\citet{lemaux18} & 150.353 & 2.338 & 4.568 & 1.5 & blind -- spectroscopy \\
\citet{sillassen22} & 150.466 & 2.636 & 3.650 & 1.7 & blind -- phot-$z$ \\
\citet{mcconachie22} & 150.117 & 2.564 & 3.380 & 0.9 & blind -- phot-$z$ \\
\citet{mcconachie22} & 149.850 & 2.427 & 3.367 & 2.6 & blind -- phot-$z$ \\
\citet{daddi22} & 149.582 & 2.603 & 3.295 & 1.3 & targeted -- radio galaxy \\
\citet{lemaux14} & 36.750 & -4.355 & 3.290 & 1.5 & blind -- spectroscopy \\
\citet{toshikawa16} & 36.150 & -4.330 & 3.130 & 1.7 & blind -- LBG \\
\citet{daddi21} & 150.346 & 2.335 & 2.910 & 3.2 & targeted -- radio galaxy \\
\citet{cucciati14} & 150.096 & 2.000 & 2.895 & 1.3 & blind -- spectroscopy \\
\hline
\end{tabular}
\\
\footnotesize{$^\text{a}$Primary method of protocluster discovery, marking the galaxy type used as
signpost (``targeted'') or to carry out the overdensity measurement (``blind'').}
\end{table*}

\bsp	
\label{lastpage}
\end{document}